\documentclass[12pt]{article}

\usepackage{amsmath,amssymb,amsthm,amscd,mathtools, graphicx, mathtext,color,wrapfig}

\usepackage[numbers,sort&compress]{natbib}
\makeatletter

\@addtoreset{equation}{section}
\makeatother

\newcommand{\beq}{\begin{equation}}
\newcommand{\eeq}{\end{equation}}

\theoremstyle{definition}

\begin{document}
\baselineskip=18pt  
\baselineskip 0.713cm

\begin{titlepage}

\setcounter{page}{0}

\renewcommand{\thefootnote}{\fnsymbol{footnote}}

\begin{flushright}
ITEP-TH XX/14
\end{flushright}

\vskip 1.0cm

\begin{center}
{\LARGE \bf
$A_n$-Triality
\\
\vskip  0.5cm
}

\vskip 0.5cm

{\large
Mina Aganagic$^{1,2}$, Nathan Haouzi$^{1}$
and Shamil Shakirov$^{1,2,3}$}
\\
\medskip

\vskip 0.5cm

{\it
$^1$Center for Theoretical Physics, University of California, Berkeley,  USA\\
$^2$Department of Mathematics, University of California, Berkeley,  USA\\
$^3$Institute for Theoretical and Experimental Physics, Moscow, Russia
}

\end{center}

\vskip 0.5cm

\centerline{{\bf Abstract}}
\medskip

$A_n$-type AGT correspondence anticipates that conformal blocks of $A_n$ Toda CFT are related to partition functions of a family of 4d ${\cal N}=2$ SCFTs. We use gauge/vortex duality to both give a precise form of the correspondence, and to prove it. Gauge/vortex duality relates the 4d theories and the 2d theories living on its vortices. Partition functions of the 2d theories on vortices provide Coulomb-gas representation of $A_n$ Toda conformal blocks with discrete internal momenta. This gives a triality of relations between the gauge theory, its vortices and the Toda CFT. We prove that $A_n$ triality holds for conformal blocks of $A_n$ Toda on a sphere with all full punctures. The lift to one higher dimensional theories, compactified on a circle of arbitrary radius, and $q$-deformation of the Toda CFT, play a key role.

\noindent\end{titlepage}
\setcounter{page}{1} 


\section{Introduction}

It is expected \cite{AGT} that there is a correspondence between a conformal block of $A_n$-type Toda CFT and a partition function of a 4d ${\cal N}=2$ gauge theory. The 4d theory, which we will denote by ${\cal T}_{4d}$, is superconformal,  belonging to a class of theories defined in \cite{G2, Gaiotto:2009hg} in terms of $n+1$ M5 branes wrapping a Riemann surface $C$. The $A_1$ case, for which the Toda CFT becomes the Liouville CFT, has been studied most extensively.

In this paper, we use gauge/vortex duality to give a precise statement of the correspondence, which we will prove, for genus zero conformal blocks. Gauge/vortex duality relates the 4d ${\cal N}=2$ gauge theory in a certain 2d background to a 2d ${\cal N}=(2,2)$ theory on its vortices  \cite{DH1, DH2, SimonsTalk}.\footnote{For earlier work see \cite{Dorey, DHT, HananyTong, HananyTong2, Shifman:2004dr}.} We will show that the correspondence between the 4d gauge theory and $A_n$ Toda conformal blocks follows from it. Since the theory on vortices plays a crucial role, the correspondence between the CFT and the 4d gauge theory is really a triality.  In fact, we will prove a more general statement, involving a one-parameter deformation of Toda CFT conformal blocks, based on replacing the $W$-algebra symmetry that governs the CFT by a $q$-deformed $W$-algebra.

\subsection{The $A_n$ Triality}

Let ${\cal T}_{5d}$ be a 5d ${\cal N}=1$ theory, compactified on a circle of radius $R$. The theory is defined by the property that its Seiberg-Witten curve agrees with, in the $R$ to zero limit, the Seiberg-Witten curve $\Sigma$ of ${\cal T}_{4d}$.  We take $\Sigma$ to be an $n+1$ fold cover of a genus zero curve $C$ with full punctures (as well as some other cases), in the language of  \cite{G2, Gaiotto:2009hg}. Let ${\cal V}_{3d}$ be a 3d ${\cal N}=2$ theory on vortices in ${\cal T}_{5d}$. We place ${\cal T}_{5d}$ in $\Omega$-background, and by restriction ${\cal V}_{3d}$ as well. Then, we prove the following:

$i.)$
The partition function of the 3d theory ${\cal V}_{3d}$ is identical to the $q$-conformal block of the $A_n$-type Toda CFT in Coulomb gas representation \cite{Dotsenko:1984nm}. Coulomb gas requires integral momenta in intermediate channels. These integers are the ranks of the 3d gauge groups.

$ii.)$ The partition function of ${\cal V}_{3d}$, computed by residues, equals the partition function of ${\cal T}_{5d}$ at certain integer values of Coulomb branch parameters, determined by the ranks of the 3d gauge groups. That the two should agree follows from gauge/vortex duality.

This generalizes the $A_1$ triality, proven in \cite{AHKS}, to general $n$. The $A_n$-triality holds for any values of the ranks $N_a$ of the 3d gauge groups, and for any choice of $q=e^{R \hbar},t=e^{-R \epsilon}$, the two parameters of $\Omega$-background.  In the large $N_a$ limit, where one keeps $N_a\epsilon $  fixed one probes arbitrary values of the Coulomb branch moduli and arbitrary conformal blocks.

In retrospect, the gauge/vortex duality implies the large $N$ duality of topological string theory  \cite{GV, DV, IH}  by choosing a self-dual $\Omega$ background, at $\epsilon+\hbar=0$.  A conjecture that large $N$ duality of topological string theory provides an explanation of \cite{AGT} was made in \cite{DVt}; we prove it here. For a review of relation to topological string large $N$ duality, see \cite{toappear}. The gauge/vortex duality should extend the BPS/CFT correspondence of \cite{Nekrasovtalk, Carlsson:2013jka} to a triality, whenever the CFT has Coulomb-gas formulation.   
\subsection{Some Finer Points}

The lift to 5d is necessary for the following reason. The ${\cal T}_{4d}$ SCFT corresponding to $C$ with all full punctures has no Lagrangian description, generically. Its 5d lift ${\cal T}_{5d}$ turns\footnote{In \cite{Benini:2009gi} theories of this type were called Sicilian gauge theories. Relation of $T_n$ theory to 5d $A_n$ quiver gauge theories was discussed recently in \cite{T1,T2, T3}.}
out to have a low energy description in terms of an $A_n$ quiver gauge theory with fundamental matter and ${\cal N}=1$ SUSY in 5d, of the type recently studied in \cite{NP, Nekrasov:2013xda}. At the same time, the 3d theory ${\cal V}_{3d}$ also has a Lagrangian description, as a 3d ${\cal N}=2$ quiver gauge theory (the quiver is the hand-saw quiver of \cite{NHS}).

For $A_{n>1}$ type Toda CFT and its $q$-deformation, conformal blocks are more complicated than in the $A_1$ case, because the
Virasoro symmetry alone is in general not enough to fix the basic building block, the sphere with 3 punctures. For example, if all three of the punctures are full, corresponding to insertions of $W$-algebra primaries, the conformal block depends on additional $n(n-1)/2$ moduli.  The relation to 3d gauge theory helps us navigate the problem. The Coulomb gas for $q$-conformal blocks of $A_n$ Toda on a sphere gives the ${\cal V}_{3d}$ partition functions. From the 3d perspective, when $n>1$, the 3d gauge theory has a fairly intricate structure of vacua and corresponding to this, the choices of contours in defining the partition functions \cite{Witten_anewlook}. These can be completely understood.  While Coulomb gas describes only conformal blocks with specializations, the contours we provide should be helpful in solving the Toda CFT for $n>1$, possibly generalizing \cite{TL}.\footnote{We are grateful for J. Teschner for discussions on this point.}

In section 2, we first review the 4d and the 5d theories, ${\cal T}_{4d}$ and ${\cal T}_{5d}$, and show that  ${\cal T}_{5d}$ has a Lagrangian description at low energies. This allows us to compute the partition function of the gauge theory. Taking the 4d limit, we get the partition function of ${\cal T}_{4d}$ as well, even though the theory itself becomes strongly coupled as we send $R$ to zero.
In section $3$ we describe the theory on vortices ${\cal V}_{3d}$ and compute its partition function. In section 4. we review Coulomb gas approach to $A_n$ Toda, and its $q$-deformation. We prove the part $i.$ of genus zero $A_n$-triality. In sections 5 we explain the physics of gauge/vortex duality and prove part $ii.$ of the $A_n$-triality.

\section{${\cal T}_{5d}$ and M5 Branes}

%

In this section, we begin by reviewing the M5 brane construction of ${\cal T}_{4d}$ and its deformation, the 5d theory ${\cal T}_{5d}$ compactified on a circle of radius $R$. Then, we show that if $C$ is a genus zero curve with {\it at least two} full punctures the 5d theory ${\cal T}_{5d}$ has a Lagrangian description. This allows us to obtain its partition function in $\Omega$-background explicitly, and, by taking the $R$ to zero limit, that of ${\cal T}_{4d}$ as well.

\subsection{M5 branes and ${\cal T}_{4d}$}

Let $\Sigma$ be the Seiberg-Witten curve of ${\cal T}_{4d}$,

\beq\label{4dcurve}
{ \Sigma}:\qquad \qquad p^{n+1} + \phi^{(2)}(z) p^{n-1} + \ldots +\phi^{(n+1)}(z)=0.
\eeq
with meromorphic one form $\lambda = pdz$. ${\Sigma}$ is $n+1$-fold cover of $C$, which we take to be a genus zero curve, with coordinate $z$. This makes $p$ a section of $T^*C$, and $\phi^{(k)}(dz)^k$ a degree $k$ differential on $C$.

The Seiberg-Witten curve encodes both the UV and the IR data of the theory. Specifying the UV data of the theory corresponds to picking a set of punctures on $C$ where the Seiberg-Witten differential $p dz$ has a pole of order one, and fixing the residues

$$(\alpha_1, \alpha_2, \ldots , \alpha_{n},\alpha_{n+1})
$$
on the $n+1$ sheets. In theories with special unitary, as opposed to unitary, gauge groups, we would subject this to the condition that $\sum_{a=1}^{n+1} \alpha_{a} =0$. The IR data, namely choosing a point on the Coulomb branch of the theory, corresponds to picking a specific $\Sigma$: this can be parameterized \cite{G2} by the set of differentials $\phi^{(k)} (dz)^k$ on ${C}$, which are holomorphic away from the punctures, and with the behavior at the punctures consistent with the UV data.

At a generic puncture all the residues are distinct. The generic puncture is a {\it full puncture}, in the terminology of \cite{G2}. In less generic cases, some residues may coincide. The punctures are thus labeled by partitions of $n+1$, or equivalently, by Young diagrams with $n+1$ boxes. Young diagram has a column of height $k$ if $k$ residues come together at the puncture. In particular, the full puncture corresponds to a diagram with a single row, of length $n+1$.

The Seiberg-Witten curve gives a fairly complicated way of encoding the theory. For conformal theories like ${\cal T}_{4d}$ there is a simpler curve that only captures the theory, as opposed to the theory and a point on the Coulomb branch -- we will call this curve the $S$-curve. To get the $S$-curve we simply take the Seiberg-Witten curve at a point of the moduli space where $\Sigma$ degenerates to  $n+1$ components:

\beq\label{4dcurve2}
{S}:\qquad \qquad\prod_{a=1}^{n+1}(p - W'_a(z)) =0
\eeq
where

$$
W'_a(z)=\sum_i {\alpha_a^{(i)}\over z- z_i}
$$
corresponding to having punctures at $z=z_i$, as well as a puncture at infinity. Unlike $\Sigma$ which is a complicated curve of high genus, the $S$-curve consists of $n+1$ copies of $C$, and encodes the UV data in a simple, manifest way. This point in the moduli space where $\Sigma$ becomes the $S$-curve the intersection of the Coulomb and Higgs branches.\footnote{As we will discuss below, for us it is natural to take the gauge groups to be of unitary type, rather than special unitary type. We can reach this point by varying moduli of the theory, as opposed to couplings, only if the gauge groups are unitary. } From the $S$-curve we can get the Seiberg-Witten curve at a generic point in the Coulomb branch moduli space by simply resolving the singularities where different components of the $S$-curve meet.\footnote{The $S$-curve is clearly not purely a UV object, since it is based on the Seiberg-Witten curve at a point in the moduli space; still, it is canonical, and can be used to succinctly encode the theory geometrically.}

\subsection{M5 branes and  ${\cal T}_{5d}$}

The 5d ${\cal N}=1$ theory  ${\cal T}_{5d}$ compactified on a circle of radius $R$,  can be defined via an M5 brane wrapping its Seiberg-Witten curve. The theory can be thought of as a deformation of ${\cal T}_{4d}$, by one parameter $R$ \cite{Lawrence:1997jr}. In particular, in the $R$ to zero limit, its Seiberg-Witten curve agrees with the Seiberg-Witten curve $\Sigma$ of ${\cal T}_{4d}$.

Like in 4d, rather than specifying the Seiberg-Witten curve of the theory, we can specify its $S$-curve. The $S$-curve is a more convenient starting point, since it encodes only the UV data of the theory, and in a manifest way. The $S$-curve of the 5d theory ${\cal T}_{5d}$, is

\beq\label{5dcurve2}
S\;\;:\qquad\qquad
\prod_{a=1}^{n+1}(e^{p} - V_a(e^x)) =0
\eeq
with the meromorphic one form equal to $\lambda = p dx$ and where

$$
V_a(e^x) =  {e^{\zeta_a }\over \prod_{i=1}^{\ell} ( 1- e^x/  f_i^{(a)}) }.
$$
The four dimensional curve is recovered by taking the $R$ to zero limit. To take the limit, one first writes

\beq\label{massb}
f_i^{(a)} = z_i\; e^{R \alpha_i^{(a)}},
\eeq
and redefines $e^p$ by replacing it with factor $e^p/\prod_{i=1}^{\ell} ( 1- e^x/  z_i)$. Then, one takes $R$ to zero keeping  $p/R$, $\zeta/R$, $z_i$ and $\alpha_{i}^{(a)}$ fixed. Finally, one defines $z=e^x$, and replaces $p$ by $pz$ to get \eqref{4dcurve2}, with its canonical one form $\lambda = pdz$. Note that one of the punctures we get is automatically placed at $z=0$.

Like in the 4d case, the $S$-curve is the Seiberg-Witten curve of ${\cal T}_{5d}$, compactified on a circle, at a point where the Higgs and the Coulomb branch of the theory meet.

\subsubsection{Classification of Punctures in 5d}

In 5d, like in 4d, the punctures are classified by the behavior of the Seiberg-Witten differential $\lambda$ near them. In 5d, the poles of $\lambda$ are logarithmic -- meaning that near a pole $\lambda\sim   \;\rm{log}(z-z_*)dz/z$ where $z=e^x$. The fact that ${\cal T}_{5d}$ reduces in the $R$ to zero limit to a 4d SCFT, implies there is an equal number of logarithmic poles on each sheet. When $n+1$ poles from different sheets come close together, at $n+1$ positions $x=x_{a,*}$ on ${C}$, it becomes natural to define the "position of the puncture" to be the average of those, $\sum_a x_{a,*}/(n+1)$. The role of residues is played by the positions of the $n+1$ punctures measured relative to this. This naturally corresponds to separating the center of mass degrees of freedom of the M5 branes from the relative ones.
Thus, in 5d, the parameters to specify at each puncture are no longer divided sharply into position of the puncture and the value of residues. Yet, we will borrow the 4d terminology, and call the "positions" of the punctures the center of mass positions in 5d, and the "residues" the positions of the punctures relative to the center of mass.

In the most generic case, corresponding to the full puncture,  we have a pole at on each of the $n+1$ sheets of the ${S}$-curve above the puncture on $C$, but at distinct values of $x$. In our terminology, all the 5d residues are distinct then. Alternatively, some of the punctures coming from different sheets may coincide at $x_{a,*}= x_{b,*}$, for some distinct $a,b$, and then the corresponding  residues coincide. In principle, it is possible to have punctures coincide coming from the same sheets of the Riemann surface, however we will disallow this. This is a realization of the "s-rule" in the language of Riemann surfaces. Riemann surfaces wrapped by the M5 branes that violate the "s-rule" are too singular to correspond to physical theories: any attempt to resolve the singularity results in breakdown of supersymmetry. Thus, as is common, we will exclude them.
In 5d, it is also possible for a number $\ell(n+1)$ of the poles to come together, where $\ell$ is an integer bigger than one. One can do so without violating the s-rule, provided some judicious choice of how punctures come together. It is easy to show that these higher $\ell$ 5d punctures become  indistinguishable from ordinary punctures, with first order poles, in the 4d limit.

We encode, as in \cite{G2}, the behavior of the $S$-curve (and the Seiberg-Witten curve $\Sigma$) near a puncture at a point $P$ on ${C}$ by a Young diagram $Y_P$. A column of height $m$ in the Young diagram $Y_P$ corresponds to having a puncture where $m$ of the residues coincide. Placing them in a column also reminds us that no two punctures corresponding to it may come from the same sheet. Finally, only those Young diagrams arise which have $n+1$ boxes -- or, exceptionally in 5d, a multiple $\ell$ of that. %

\subsection{${\cal T}_{5d}$ is a 5d Quiver Gauge Theory}

A sign of usefulness of going to five dimensions is that ${\cal T}_{5d}$ has a Lagrangian description as a quiver gauge theory of $A_n$ type.\footnote{This holds at intermediate energies. In the far UV, the theory is a strongly coupled 5d ${\cal N}=1$ SCFT of $A_n$ type, which can be defined using string theory.} A  sufficient condition for this is that the parameters $\zeta_a$ entering the 5d $S$-curve are generic, since $\zeta_a$ determine the 5d gauge couplings. For the partition function to make sense as a power series in instanton contributions, all we need is $|e^{\zeta_a}|<1$. This corresponds to a theory of $n+1$ M5 branes on a genus zero curve $C$ with two full punctures, at $z=0$ and $z=\infty$, and where the rest of the punctures can be {\it arbitrary}. (The simplest case turns out to be the most generic one -- where all the punctures on $C$ are full.) Five dimensional gauge theories of this type, compactified on a circle, were studied in \cite{N2, NO, NP}, and elsewhere. Thus, introducing a circle allows for a much larger class of theories with a Lagrangian description. The quiver theory we will end up with is {\it not} the linear quiver theory of \cite{G2, W7}.

\subsubsection{A Class of 5d $A_n$-Type Quiver Gauge Theories}

\begin{figure}[h!]
\begin{center}
\emph{}
\hspace{-7ex}
\includegraphics[width=0.5\textwidth]{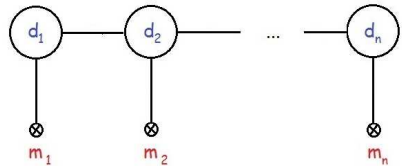}
\end{center}
\caption{Quiver for gauge theories of type $A_n$.}
\label{GaugeQuiver}
\end{figure}

The 5d quiver gauge theory we need is based on the Dynkin diagram of the $A_n$ Lie algebra. We will label the nodes of the Dynkin diagram with integers $a=1, \ldots n$, as in the figure \ref{GaugeQuiver}. For each node we associate a unitary gauge group factor, so that the node labeled by $a$ corresponds to the gauge group $U(d_a)$. The gauge symmetry group ${G}_{5d}$ is

\beq\label{5dg}
{G}_{5d}= \otimes_{a=1}^n U(d_a).
\eeq
There is a bifundamental hypermultiplet, transforming in $(d_a, {\overline d_b})$ for every pair of nodes in the Dynkin diagram with a link between them. The theory also has a set of $m_a$ matter fields in the fundamental representation of $U(d_a)$, so that the flavor symmetry group is

\beq\label{5df}
F_{5d} = \otimes_{a=1}^n U(m_a).
\eeq
At each node, the ranks $d_a$ satisfy

\beq\label{conf}
\sum_b C_{ab}\; d_b= m_a,
\eeq
where we set $d_0=d_{n+1}=0$. The matrix $C_{ab}$ is the Cartan matrix of the Lie algebra, $C_{ab} = 2 \delta_{ab} - \delta_{a, a+1} -\delta_{a, a-1}.$

The Cartan matrix is invertible, so the ranks of the gauge groups of the 5d quiver are uniquely determined from the flavor symmetry group $F_{5d} = \prod_{a=1}^nU(m_{a})$. However, not every choice of $F_{5d}$ is allowed, since we need to require $n_a$ to be integers, $n_a \in {\mathbb Z}_{\geq 0}$. The choices of the integers $m_a$ for which \eqref{conf} has solutions over integers can be shown to correspond to Young diagrams $Y_F$ with a total of ${\ell}(n+1)$ boxes, where $m_a$ is the number of columns of height $a$ in $Y_F$. For every such flavor Young diagram
$$Y_F
$$
we get a 5d $A_n$ quiver theories satisfying \eqref{conf}. In addition, the 5d gauge theories can have Chern-Simons terms. The 5d Chern-Simons levels $k_a$ of ${\cal T}_{5d}$ are $k_a= d_a-d_{a+1}$, for the $a$'th gauge group.

%
%
%

\begin{figure}[h!]
\begin{center}
\emph{}
\hspace{-7ex}
\includegraphics[width=0.4\textwidth]{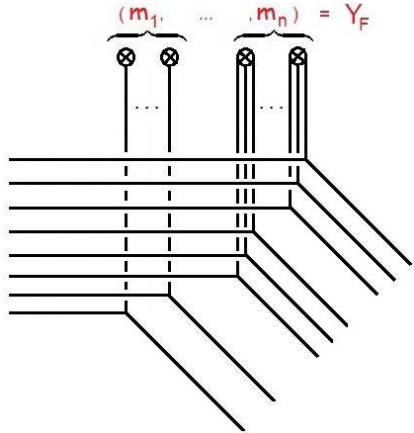}
\end{center}
\caption{A \emph{web diagram} for 5d $A_n$ quiver theories, see section 2.5.}
\label{GaugeWeb}
\end{figure}

\subsubsection{From $S$-curve To $A_n$-Quiver Gauge Theory}

The 5d theory ${\cal T}_{5d}$ is a priori defined by the $S$-curve. We will associate to every $S$-curve in \eqref{5dcurve2} a 5d $A_n$ quiver gauge theory in Fig. \ref{GaugeQuiver}.
The $S$-curve is the Seiberg-Witten curve of the quiver theory, at the point in the moduli space where the Coulomb and Higgs branch meet. Thus, the quiver gauge theory provides a Lagrangian description of the 5d theory ${\cal T}_{5d}$. Let us first state the correspondence, and then explain how to obtain it.

The $S$-curve is an $n+1$ fold covering of $C$ with punctures at points $P_i$. For each such puncture at $z\neq 0, \infty$, we associated a Young diagram
$$
Y_{P_i}
$$
with ${\ell}_{P_i}(n+1)$ boxes labeling how sheets of $S$ come together over the puncture. The simplest choice, the full puncture, where the residues at the puncture were all distinct, corresponded to the Young diagram with one row and $n+1$ boxes, so ${\ell}_P=1$ in this case. When $a$ residues coincide at $P_i$, $Y_{P_i}$ has a column of height $a$. We discussed meaning of the residues and the positions of punctures for the 5d curves in the previous subsection.

The theory ${\cal T}_{5d}$ associated with the $S$-curve is the quiver gauge theory in figure \ref{GaugeQuiver} with the ranks determined by  the flavor Young diagram $Y_F$ which is a sum of the Young diagrams associated to the punctures on $S$

$$
Y_F = \oplus_i Y_{P_i}.
$$
We define the sum of two Young diagrams $Y$ and $Y'$ with $m_a$ and $m_a'$ rows of height $a$ to be a diagram $Y\oplus Y'$  with $m_a + m_a'$ rows of height $a$.

To establish this, we proceed as follows. First, we will deform the $S$-curve corresponding to going to the generic point of its moduli keeping the asymptotics fixed, resulting in an irreducible curve $\Sigma$.  On the one hand, we show that the resulting $\Sigma$ is the Seiberg-Witten curve of a the 5d gauge theory in Fig.\ref{GaugeQuiver} with flavor symmetry encoded by $Y_F$. On the other hand, per definition, deforming $S$-curve to $\Sigma$ corresponds to going onto the generic point of the Coulomb branch of ${\cal T}_{5d}$. This will complete what we want to show.

The fact that the sum of the Young diagrams at the punctures equals the Young diagram $Y_F$ implies that all together, on the $S$-curve in \eqref{5dcurve2} there are $m_b$ distinct values of $x$, where $b$ punctures come together. In terms of the functions $V_a(e^x)$ that enter the $S$-curve, this means that there exist $n$ functions $Q_{m_{b}}(e^x)$'s which are polynomials in $e^x$ with zeros where $V_a$'s have poles,  such that

$$
\prod_{a=1}^{n+1}V_a \times Q_{m_1} \Bigl(Q_{m_2}\Bigr)^2 \Bigl(Q_{m_3}\Bigr)^3\ldots \Bigl(Q_{m_{n}}\Bigr)^n  = 1.
$$
and where

$$Q_{m_b}(e^x) = \prod_{i=1}^{m_b}(1- e^x/{f}_{b; i}),
$$
The zeros of $Q_{m_b}$ are assumed to be all distinct. Multiplying the $S$-curve by $\prod_{b=1}^n \Bigl(Q_{m_b}\Bigr)^b$,  and resolving singularities
we can rewrite it as

\beq\label{5dsw}
\Sigma: \qquad \prod_{b=1}^n \Bigl(Q_{m_b}\Bigr)^b t^{n+1} + \prod_{b=2}^n \Bigl(Q_{m_b}\Bigr)^{b-1} G_{d_1} t^n +\ldots +  G_{d_n} t+1=0
\eeq
where $t=e^p$, and the coefficient of $t^{n-k+1}$ equals $\prod_{b=k+1 }^n \Bigl(Q_{m_b}\Bigr)^{b-k} G_{d_k}(e^x)$. Simply rewriting the $S$-curve in this form, results in  polynomials $G_{d_k}(e^x)$ with coefficients fixed by \eqref{5dcurve2}. Deforming away from this to $G_{d_k}(e^x)$ a generic degree $d_k$ polynomial in $e^x$, we get the curve $\Sigma$ in  \eqref{5dsw}.

The claim is that the curve $\Sigma$ is the Seiberg-Witten curve of the 5d ${\cal N}=1$ quiver gauge theory in Fig. \ref{GaugeQuiver}, compactified on a circle, with Seiberg-Witten one form $\lambda = x dt/t$. To see this, we will show that there is {\it another} four dimensional limit, in which the 5d Seiberg-Witten curve in \eqref{5dsw} becomes the Seiberg-Witten curve of the 4d ${\cal N}=2$ gauge theory with the quiver given in Fig.\ref{GaugeQuiver}. (Note that we are not claiming this 4d theory is the same as ${\cal T}_{4d}$, or even dual to it -- different four dimensional limits result in different 4d ${\cal N}=2$ gauge theories. The 4d limit here merely aids in identifying the Lagrangian description of ${\cal T}_{5d}$.) To take the four dimensional limit, we need to reinstate $R$, the radius of the compactification circle, and take the limit when $R$ goes to zero. We do so by writing $f_{i}^{(a)}$ as $f_i^{(a)} = e^{R \mu_i^{(a)}}$, and keeping $x/R$, $e^{p R}$, $e^{\zeta_a R}$ and the $\mu$'s fixed in the limit. The effect of this is that the 4d curve has the same form as \eqref{5dsw}, but with $Q$ and $G$ replaced by polynomials of the same degree, but in $x$, rather than $e^x$. From the results of \cite{W7, G2}, this is the Seiberg-Witten curve of the $A_n$ quiver theory in Fig. \ref{GaugeQuiver}, as claimed.

The $S$-curve \eqref{5dcurve2} encodes the coupling constants and the mass parameters of ${\cal T}_{5d}$, in addition to the flavor symmetry group and the gauge group. The mass parameters are associated to the positions of the punctures at $z\neq 0, \infty$, and the values of the residues there.  The residues at the $z=0$ puncture are related to the values of the gauge couplings in the $n$ gauge group factors in $G_{5d}$. The values of the residues at $z=\infty$ are not independent parameters.
\subsubsection{4d vs. 5d}

While the ${\cal T}_{5d}$ has a Lagrangian description as a 5d $A_n$ quiver gauge theory, ${\cal T}_{4d}$ does not have a Lagrangian description in general. This is because taking the four dimensional limit to get \eqref{4dcurve2}, we also scale the gauge couplings to infinity, by necessity.
The exception is the case when the 5d theory we start out with has {\it two} inequivalent Lagrangian descriptions, related by spectral duality. The phenomenon of spectral duality was first noted in \cite{Katz:1997eq}, and was revisited later in \cite{Mironov:2012uh, Mironov:2013xva}. Then taking the $R$ to zero limit, the 4d theory we end up with can have a Lagrangian description -- based on the spectral dual 5d description.
When we describe the theory in terms of M5 branes on a Riemann surface with Seiberg-Witten one form $\lambda$ with $d\lambda = dp \wedge dx$, the spectral duality corresponds to the exchange of the roles of $p$ and $x$, accompanied by the exchange of Coulomb branch moduli and mass parameters, with the gauge couplings.\footnote{In 5d, $x$ and $p$ are on the same footing a priori, as the curve lives in ${\mathbb C}^*\times {\mathbb C}^*$, with coordinates $e^x$ and $e^p$. } This can end up relating two a priori distinct gauge theory descriptions of the same 5d theory compactified on a circle.  The 5d theory ${\cal T}_{5d}$ has an $A_n$ quiver gauge theory description as long as $C$ has {\it at least} $2$ full punctures and a number of other arbitrary punctures. It will have another, spectral dual Lagrangian description, if there are exactly two other punctures, at $P_1$ and $P_2$ on $C$, with $\ell_{P_{1,2}}=1$.  In this case, we can trade the $2$ full punctures for $n$ simple punctures, and we end up with a linear quiver description of \cite{G2}.

\subsection{Partition Function of ${\cal T}_{5d}$}

Given the Lagrangian description of ${\cal T}_{5d}$ as an $A_n$-quiver gauge theory, we can compute the partition function of the theory in 5d $\Omega$-background. The background is defined as a twisted product

\beq\label{back}
({\mathbb C}\times {\mathbb C}\times S^1)_{q,t},
\eeq
where as, one goes around the $S^1$, one rotates the two complex planes by $q = \exp(R \epsilon_1)$ and $t^{-1}=\exp(R \epsilon_2)$. These are paired together with the 5d $U(1)_R\subset SU(2)_R$ symmetry twist by $t q^{-1}$, to preserve supersymmetry. The 5d gauge theory partition function in this background is the trace

\beq\label{5dtrace}
{\cal Z}_{{\cal T}_{5d}}(\Sigma)={\rm Tr} (-1)^F {\bf g}_{5d},
\eeq
corresponding to looping around the circle in \eqref{back}. Insertion of $(-1)^F$ turns the partition function of the theory to a supersymmetric partition function. One imposes periodic identifications with a twist by ${\bf g}$ where ${\bf g}$ is a product of simultaneous rotations: the space-time rotations by $q$ and $t^{-1}$, the $R$-symmetry twist,  flavor symmetry rotations $f_{i, \pm} = \exp(R m_{i, \pm})$, and gauge rotation by $e_i = \exp(R a_i)$ for the $i$'th $U(1)$ factor. The latter has the same effect as turning on a Coulomb-branch modulus $a_i$ (see \cite{Nekrasov:2004vw, V:3} for a review). The partition function of ${\cal T}_{5d}$ in this background is computed in \cite{N2}, using localization. The partition function is a sum

\beq\label{bN}
{\cal Z}_{{\cal T}_{5d}}(\Sigma) = r_{5d} \;\sum_{\{R\}} \ e^{\zeta \cdot  R} \ I_{{\cal T}_{5d}; \{R\}}
\eeq
over a collection $\{R\}$ of 2d partitions:

$$
\{R\} = \{{R}_{a,i}\}^{a=1, \ldots n}_{i=1, \ldots d_{a}}
$$
We get a Young diagram for each $U(1)$ factor in $G_{5d}$. There are $n$ nodes of the quiver, with $U(d_{a})$ gauge group at the $a$'th node, so we get $d_{a}$ Young diagrams for this node. The summand is a product of factors.

\beq\label{5dp}
I_{{\cal T}_{5d}; \{R\}}= \prod_{a=1}^n z_{V_a, {\vec R}^{a}} \; z_{H_a, {\vec R}^{a}}\; z_{CS, {\vec R}^{a}} \; \cdot \prod^n_{a,b=1}
z_{H^{a,b}, {\vec R}^{a},  {\vec R}^{a}}
\eeq
The $a$-th node of the Dynkin diagram contributes

$$
z_{V_a, {\vec R}^{a}}= \prod_{1\leq i,j\leq d_{a}}[N_{R_{a,i} R_{a,j}}(e_{a,i}/e_{a,j})]^{-1}.
$$
Here $e_{a,i}$ encode the $d_a$ Coulomb branch parameters of $U(d_a)$ gauge group, and $N_{RP}(Q)$ is the Nekrasov function, defined below.
The $m_a$ fundamental hypermultiplets at this node contribute:

$$
z_{H_a, {\vec R}^{a}} = \prod_{1\leq i\leq d_a} \prod_{1\leq j \leq m_a} N_{\varnothing R_{a,i}}( v f_{a, j}/e_{a, i}),
$$
Here $f_{a, j}$ encode the masses of the hypermultiplets. The contribution of 5d Chern-Simons terms for this node reads
$$
z_{CS, {\vec R}^{a}} = \prod\limits_{1\leq i\leq d_a} \big(T_{R_{a,i}}\big)^{d_a - d_{a+1}}
$$
Here $T_{R}$ is the framing factor, defined below. For every pair of nodes $a,b$ connected by an (oriented) arrow in the Dynkin graph, one gets a factor

$$
z_{H^{a,b}, {\vec R}^{a},  {\vec R}^{b}}= \prod_{1\leq i\leq d_{a}}\prod_{1\leq j\leq d_{b}}[N_{R_{a,i} R_{b,j}}(e_{a,i}/e_{b,j})]^{I_{a,b}}.
$$
where $I_{a,b}$ is the incidence matrix, equal to either $1$ or $0$, depending on whether, in the Dynkin graph, there is an arrow starting at the $a$'th node and ending on the $b$'th. The basic building block of the partition function is the Nekrasov function
\begin{align*}
N_{RP}(Q) = \prod\limits_{i = 1}^{\infty} \prod\limits_{j = 1}^{\infty}
\dfrac{\varphi\big( Q q^{R_i-P_j} t^{j - i + 1} \big)}{\varphi\big( Q q^{R_i-P_j} t^{j - i} \big)} \
\dfrac{\varphi\big( Q t^{j - i} \big)}{\varphi\big( Q t^{j - i + 1} \big)}.
\end{align*}
with $\varphi(x) = \prod\limits_{n=0}^{\infty}(1-q^n x)$ being the quantum dilogarithm we previously introduced.
$ T_R =(-1)^{|R|} q^{\Arrowvert R\Arrowvert/2}t^{-\Arrowvert R^t\Arrowvert/2}$, and $v = {(q/t)^{1/2}}$ as before (we use the conventions of \cite{Awata:2008ed}).

The partition function depends on two sets of parameters. The moduli of the $S$-curve correspond to the choice of the theory, ${\cal T}_{5d}$: these are the inverse gauge couplings, $\zeta_a$ and the mass parameters $f_{(a), \alpha}$. The gauge couplings keep track of the total instanton charge, via the combination

$$\zeta \cdot R = \sum_{a=1}^n  \sum_{i=1}^{d_a}\;\zeta_a\; |R_{a,i}|.
$$
There are $m_a$ hypermultiplet masses for the node $a$, encoded in $f_{a}$'s. The hypermultiplets at $a$-th node come from points on $C$ where $a$ residues of the $S$-curve come together. The rest of the parameters in the partition function, the $e_{a}$'s, label a choice of a point of on the Coulomb branch of
the theory ${\cal T}_{5d}$. The normalization factor $r_{5d}$ in \eqref{5dp} contains the perturbative and the one loop contributions to the partition function.

\subsection{Other Realizations of ${\cal T}_{5d}$ in String Theory}

There are several other useful realizations of ${\cal T}_{5d}$ in string theory. Starting with an M5 brane wrapping the $S$-curve, corresponding to ${\cal T}_{5d}$ compactified on a circle, one can use string dualities to obtain\footnote{This follows by compactifying M-theory with an M5 brane on the $S$-curve on a $T^2$ transverse to the M5 brane. Since the $T^2$ is transverse to the branes, it does not change the low energy physics. By shrinking one of the cycles of the $T^2$ first, we go to down to IIA string with an NS5 brane wrapping the $S$-curve. T-dualizing on the remaining compact transverse circle, we obtain IIB on $Y_{S}$.}
IIB string on a Calabi-Yau 3-fold $Y_S$

\beq\label{5dcurveCY}
Y_S\;\;:\qquad\qquad
\prod_{a=1}^{n+1}(e^{p} - V_a(e^x)) =uv.
\eeq
with holomorphic $(3,0)$ form $du dx dp/u$ -- which is defined by the same data as the $S$-curve. It turns out that a Calabi-Yau of this form has a mirror $X_S$, and mirror symmetry implies that IIB string on $Y_S$ is the same as IIA string on $X_S$. The latter theory is the same as M-theory on $X_S\times S^1$. The five dimensional theory ${\cal T}_{5d}$ is the 5d ${\cal N}=1$ theory that arises in the low energy limit of M-theory on $X_S$.

Another way to obtain ${\cal T}_{5d}$ is via $(p,q)$ five brane webs, using M-theory on $T^2$/type IIB on $S^1$ duality \cite{Leung:1997tw}. The mirror Calabi-Yau $X_S$ is  toric, meaning that it is a $T^2$ fibration, with fibers that degenerate over a four dimensional base. M-theory/IIB duality relates loci where $(p,q)$ cycles of the $T^2$ degenerate to locations of $(p,q)$ 5-branes in IIB. To the $S$-curve we can thus associate a $(p,q)$ 5-brane web. At low energies, the web gives rise to the 5d ${\cal N}=1$ theory which we called ${\cal T}_{5d}$ \cite{Aharony:1997bh}. This explains why M5 branes on the 5d $S$-curve \eqref{5dcurve2} describe a five dimensional theory on a circle, instead of a purely four dimensional theory.

The $(p,q)$ web diagram associated to 5d quiver theories ${\cal T}_{5d}$ given earlier can easily be worked out, starting from the $S$-curves. The poles of the $S$-curves located at finite $x$ all correspond to D5 branes, or $(1,0)$ branes. Coincident poles on different sheets correspond to D5 branes colliding together. These boundary conditions can be represented by D5 branes ending on the D7 branes. Pole of the $S$-curves at $e^x=0$ correspond to positions of NS5 branes or $(0,1)$ branes. We will take these to be generic, corresponding to a full puncture there. From this data, the $(p,q)$ 5-brane charge conservation at each vertex, and the BPS condition, one can determine the rest of the web diagram.\footnote{More succinctly, the web diagram is the tropical limit of the $S$-curve.}

In what follows, we will use these three pictures interchangeably, as they each make different aspects of the theory easier to capture.

\subsubsection{Examples of 5d Quivers, $S$-curves and Web Diagrams}

We give some examples of the 5d $A_4$ quiver gauge theories and the corresponding $C$-curves with punctures, the $S$-curves and the web diagrams. For simplicity, we restrict to the cases with three and four punctures.

\begin{figure}[t!]
\begin{center}
\includegraphics[width=0.4\textwidth]{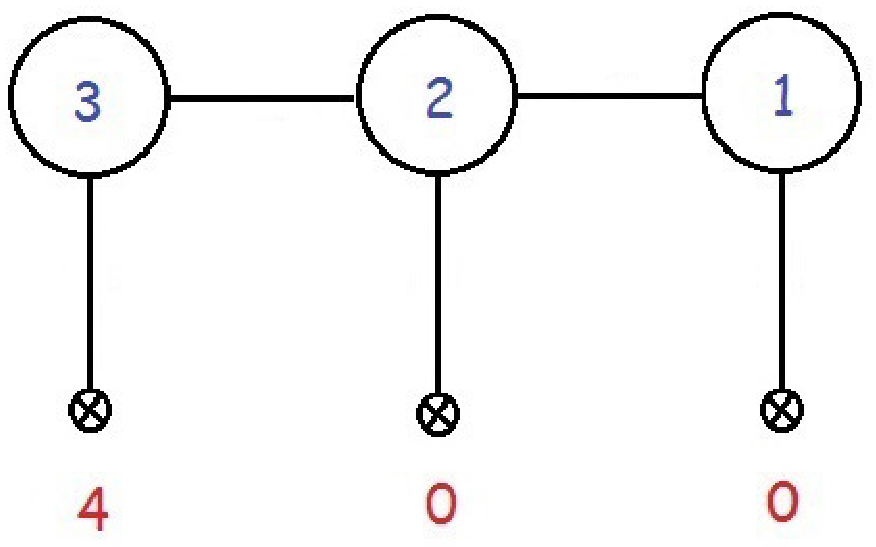}
\end{center}
\begin{center}
\includegraphics[width=0.4\textwidth]{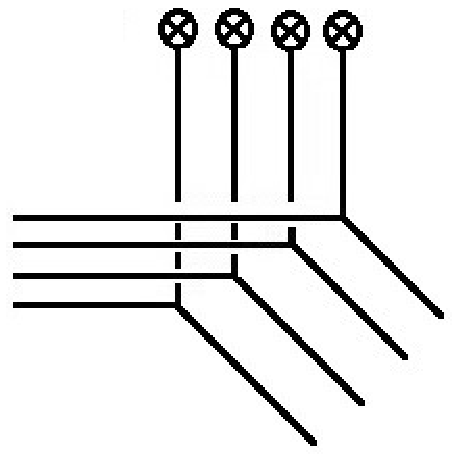} \ \ \ \ \
\includegraphics[width=0.4\textwidth]{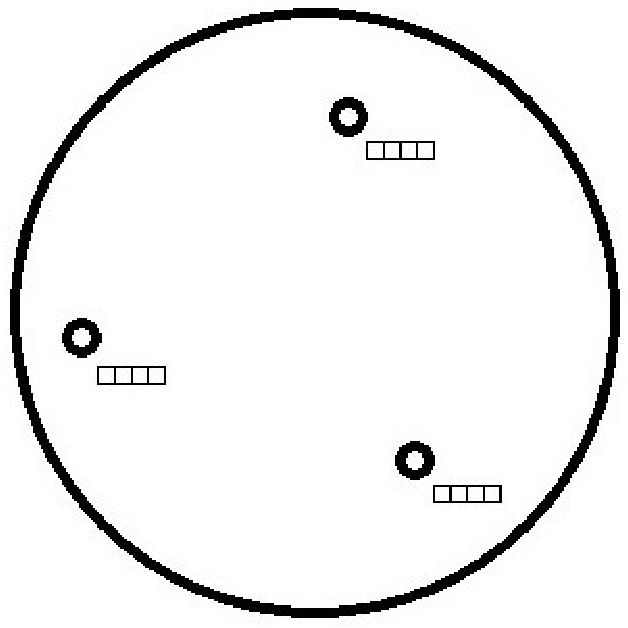}
\end{center}
\caption{The $A_4$ quiver gauge theory with ${\vec m} = (4,0,0)$, the corresponding web diagram, and the $C$ curve, which is a sphere with three full punctures.}
\label{400figure}
\end{figure}

The simplest example of this class of theories is a theory with ${\vec m} = (4,0,0)$, i.e. with 4 hypermultiplets transforming in the fundamental representation of the leftmost gauge group, as shown on the left part of Fig. \ref{400figure}. Such a theory is geometrically engineered with the help of the Calabi-Yau defined by the web diagram shown in the center of Fig. \ref{400figure}. The ranks of gauge groups ($3$, $2$, and $1$) can be read off the diagram by counting the number of vertical lines between the given adjacent horizontal lines, and the four external vertical lines correspond to the four fundamentals. From this it follows that an $S$-curve, associated to this gauge theory, is
$$
\prod\limits_{a = 1}^{4} \left( e^{p} - \dfrac{e^{\zeta_a}}{1 - e^{x}/f^{(a)}_1} \right) = 0
$$
This is a four-fold cover of the curve $C$ -- sphere with three punctures -- with the four factors corresponding to four individual sheets. Since the positions of poles at each of the punctures are generic, all three punctures are full, as shown on the right part of Fig. \ref{400figure}.

\begin{figure}[h!]
\begin{center}
\includegraphics[width=0.4\textwidth]{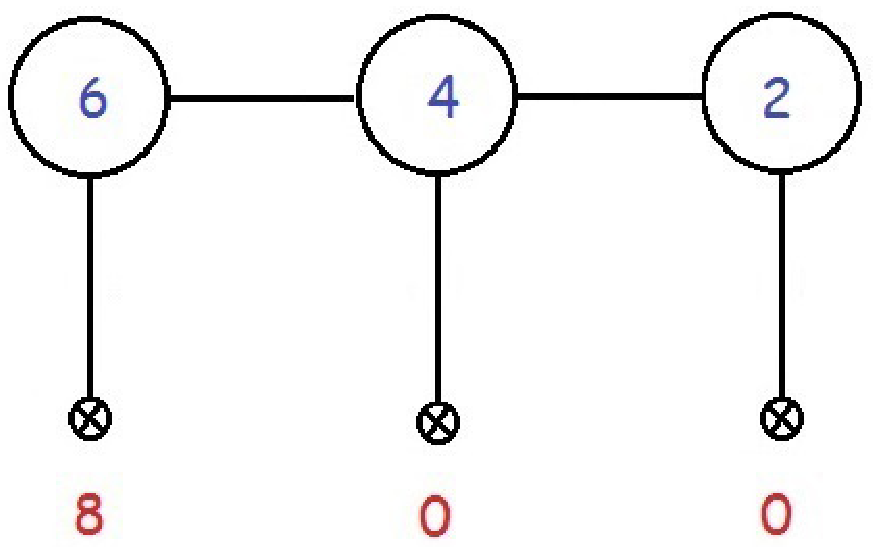}
\end{center}
\begin{center}
\includegraphics[width=0.4\textwidth]{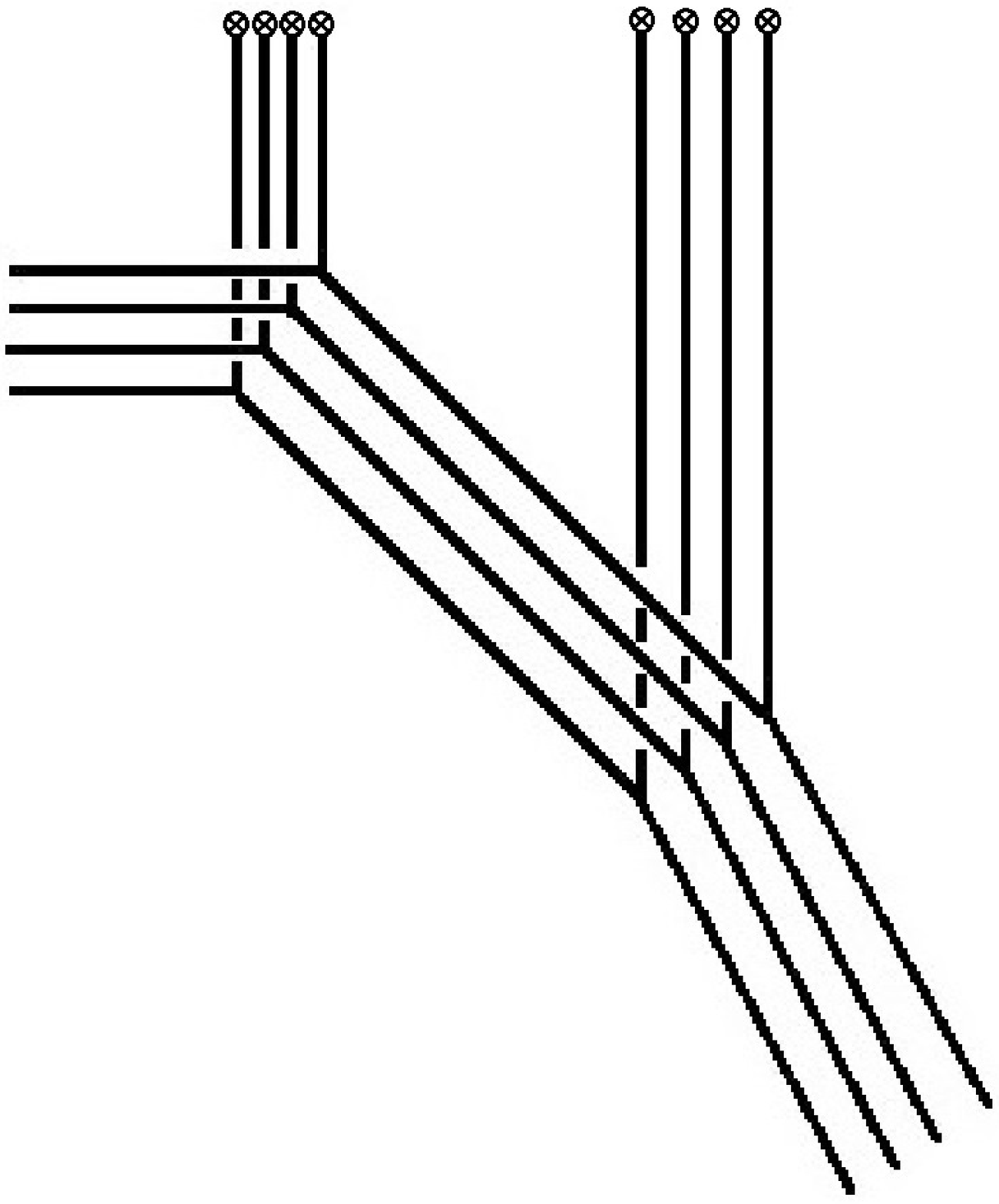} \ \ \ \ \
\includegraphics[width=0.4\textwidth]{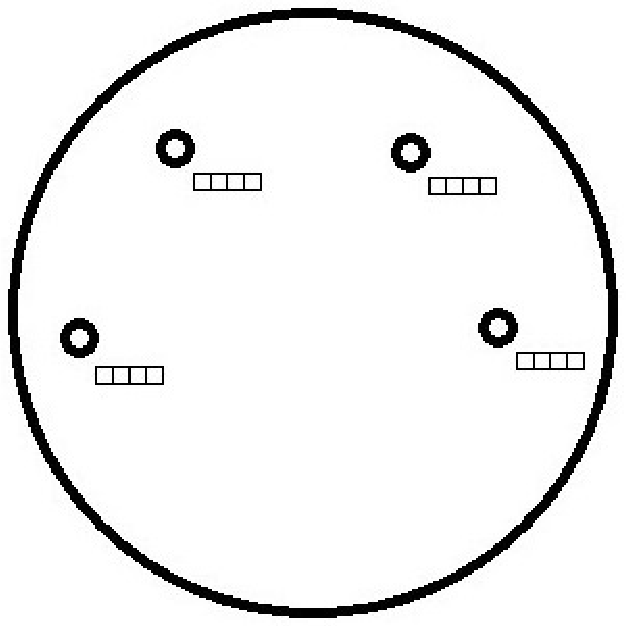}
\end{center}
\caption{The $A_4$ quiver gauge theory with ${\vec m} = (8,0,0)$, the corresponding web diagram, and the $C$ curve, which is a sphere with four full punctures.}
\label{800figure}
\end{figure}

To illustrate a possibility of having more punctures, one could consider a theory with ${\vec m} = (8,0,0)$, i.e. with 8 fundamentals of the leftmost gauge group, as shown on the left part of Fig. \ref{800figure}. Such a theory is geometrically engineered with the help of the Calabi-Yau defined by the web diagram shown in the center of Fig. \ref{800figure}. The ranks of gauge groups ($6$, $4$, and $2$) can be read of the diagram by counting the number of vertical lines between given adjacent horiszntal lines, and the eight external vertical lines correspond to the eight fundamentals. Hence the $S$-curve is
$$
\prod\limits_{a = 1}^{4} \left( e^{p} - \dfrac{e^{\zeta_a}}{\big(1 - e^{x}/f^{(a)}_1\big)\big(1 - e^{x}/f^{(a)}_2\big)} \right) = 0
$$
Again, this is a four-fold cover of the curve $C$ -- sphere with four punctures. Since the positions of poles at each of the punctures are generic, all four punctures are full, as shown on the right part of Fig. \ref{800figure}.

\begin{figure}[h!]
\begin{center}
\includegraphics[width=0.35\textwidth]{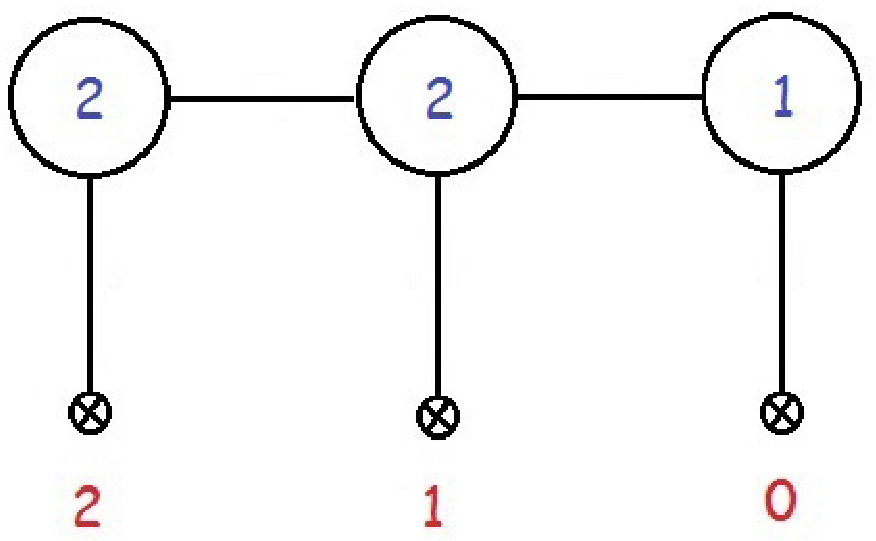}
\end{center}
\begin{center}
\includegraphics[width=0.35\textwidth]{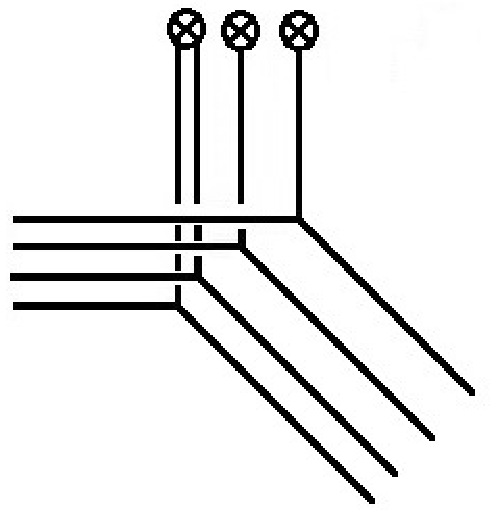} \ \ \ \ \
\includegraphics[width=0.35\textwidth]{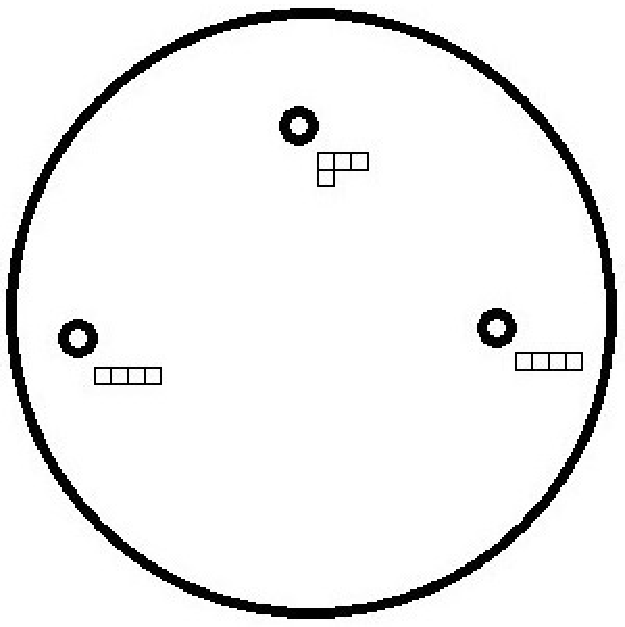}
\end{center}
\caption{The $A_4$ quiver gauge theory with ${\vec m} = (2,1,0)$, the corresponding web diagram, and the $C$ curve, which is a sphere with two full punctures and one puncture of type $[1^2 2^1]$.}
\label{210figure}
\end{figure}

To illustrate a possibility of having complicated (non-full) punctures, one could consider a theory with ${\vec m} = (2,1,0)$, i.e. with 2 fundamentals of the leftmost gauge group and one fundamental of the middle gauge group, as shown on the left part of Fig. \ref{210figure}. Because of the middle fundamental, such a theory cannot be immediately engineered with the help of a toric Calabi-Yau. However, it can still be described by a limit of a toric Calabi-Yau, namely, the web diagram shown in the center of Fig. \ref{210figure}. The black dot between two of the vertical lines stands for the limit, in which the positions of these lines are coincident, classically\footnote{The geometry of the curve captures, per definition, the moduli at $q=1=t$}. It follows that an $S$-curve is

$$
\prod\limits_{a = 1}^{4} \left( e^{p} - \dfrac{e^{\zeta_a}}{\big(1 - e^{x}/f^{(a)}\big)} \right) = 0
$$
whith $f^{(3)}=f^{(4)}$.
This is a four-fold cover of the curve $C$ -- a sphere with two full and one complicated puncture, the type of which is determined by the degeneracy of the four lines $[1^2 2^1]$ (two generic, and a merging pair).

\pagebreak

\section{${\cal V}_{3d}$ and Vortices in ${\cal T}_{5d}$}

Related to the $S$-curve and ${\cal T}_{5d}$ is another theory -- ${\cal V}_{3d}$, a three dimensional ${\cal N}=2$  gauge theory. We will first describe the theory and then explain the connection to ${\cal T}_{5d}$ and the $S$-curve.

\begin{figure}[h!]
\begin{center}
\includegraphics[width=0.6\textwidth]{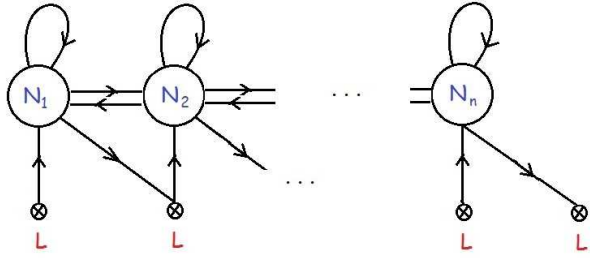}
\end{center}
\vspace{-3ex}
\caption{The handsaw quiver.}
\vspace{-3ex}
\label{HandsawFigure}
\end{figure}

\subsection{A Class of 3d Quiver Gauge Theories}

${\cal V}_{3d}$ is an $A_n$-type quiver theory, with gauge group

\beq\label{3dg}
G_{3d} = \otimes_{a=1}^n U(N_a).
\eeq
The Dynkin diagram of $A_n$ underlies the quiver diagram.
To a node labeled by $a$ in the $A_n$-Dynkin diagram, we associate the gauge group factor $U(N_a)$. There is a flavor in bifundamental representation $(N_a, {\overline N}_{b})$ for a pair of nodes that are linked in the Dynkin diagram. At each node, there is a chiral field in the adjoint representation of the $U(N_a)$ gauge group. Additional matter, dresses up the $A_n$ Dynkin diagram: there are ${\cal \ell}$ flavors in fundamental representation, with flavor symmetry rotating the chirals as in the Fig. \ref{HandsawFigure}. (A flavor in representation $R$ of the gauge group consists of a pair of chiral multiplets transforming in $R\oplus {\overline R}$.) There are two types of superpotential terms: the first couples the bifundamentals and the adjoint chiral fields at each node. The superpotential is determined by requiring  the sector of the theory obtained by forgetting the fundamental flavors to have ${\cal N}=4$, $d=3$ supersymmetry. The second type of term is associated with the "teeth" of the hand-saw. It couples together a pair of chiral fields in fundamental representation at neighboring nodes with bifundamental chirals between them. The more detailed aspects of the superpotential, such as the coefficients, do not affect the rest. The 3d Chern-Simons terms are set to zero. The quiver appeared earlier in the work of Nakajima \cite{NHS}; we will come back to this later.

The flavor symmetry group of the theory is

\beq\label{3df}
F_{3d}\times U(1)_t
\eeq
where $F_{3d} =  \otimes_{a=0}^{n} U({\ell})$ acts on the matter fields according to a the quiver diagram in Fig. \ref{HandsawFigure}. The $U(1)_t$ symmetry has its origin in the R-symmetry of the ${\cal N}=4$ theory at ${\ell =0}$, before we couple the fundamental matter to the theory. Under this symmetry, the adjoint chiral fields which originate from ${\cal N}=4$ vector multiplet have charge $1$ while the pair of chiral multiplets in the bifundamental hypermultiplet have charge $-{1/2}$, so that the superpotential coupling them is invariant. The $U(1)_t$ symmetry action on the fundamental flavors is fixed by the second type of superpotential term, the one related to teeth. There is a second $U(1)$ symmetry which which also originates from the $R$ symmetry of the ${\cal N}=4$ theory, which becomes the R-symmetry of the 3d ${\cal N}=2$ theory.

The flavor symmetry of the theory allows us to give masses to some of the matter fields. This can be done by weakly gauging the global $F_{3d}\times U(1)_t$ symmetry, giving expectation values to the scalars in the corresponding  vector multiplets, and then setting their gauge coupling back to zero. Upon compactifying the 3d theory on a circle, the theory has ${\cal N}=(2,2)$ supersymmetry, the scalars in the vector multiplets get complexified, and the corresponding complex mass parameters are the twisted masses. As we will see, these are the parameters that appear in \eqref{5dcurve2}.

\subsection{The $S$-curve and ${\cal V}_{3d}$}

The $S$-curve \eqref{5dcurve2} ends up encoding key aspects of physics of ${\cal V}_{3d}$, at low energies. On the Coulomb branch, the gauge group ${G}_{3d}$ is broken to its maximal abelian subgroup by giving expectation values to the scalars $x_{(a)}$ in the 3d vector multiplets. Integrating out all the charged matter fields and the $W$ bosons generates, at one loop, the effective twisted superpotential. The structure of the twisted superpotential in theories of this kind is reviewed recently in \cite{NS, NS1}.  In a three dimensional gauge theory on a circle (we set the radius of the circle, $R$, to $1$ in most formulas),  a chiral multiplet of twisted mass $m$, transforming in representation $Q$ of the gauge group contributes to the twisted superpotential by \cite{OV, NS1}

$$
\Delta {\cal W} =  {\pi R \over 2} {\rm Tr}_Q (x+m)^2 +{1\over 2\pi R}{\rm Tr}_Q \;{\rm Li}_2(-2\pi R(x+m)).
$$
where ${\rm Li}_2(x)= \sum_{n=1}^{\infty}{1\over n^2} e^{-nx}$ is the classical dilogarithm.
%
%
%
%
%
One can show \cite{NS1} that the contribution of $W$-bosons is equal and opposite to that of a {\it massless} adjoint chiral field. The twisted superpotential is sensitive to the ordinary superpotential only in so much as the latter restricts the flavor symmetry of the theory, and the twisted masses we are allowed to turn on.

Consider turning on the twisted masses associated with $F_{3d}$ only, and setting the twisted mass corresponding to $U(1)_t$ to zero (this sets $t=1$).\footnote{This is natural as $t$ is not a geometric parameter, visible in the curves.}  One thus specifies $\ell$ complex parameters $x_{*,i}^{(a)}$ for each of the $n+1$ flavor groups (here $a$ takes $n+1$ values and labels the flavor group, and $i$ runs from $1$ to ${\ell}$). In this case, the contributions to the superpotential of the $W$-bosons, the adjoint scalars, and the bifundamentals all cancel, and only the fundamental and anti-fundamental matter contributes. Hence, the effective superpotential is of the form

\beq\label{3dW}
{\cal W} = \sum_{a=1}^n {\rm Tr}_{U(N_a)} {\cal W}_a(x^{(a)})
\eeq
where the trace is the trace in fundamental representation, and ${\cal W}_a$ is the superpotential at node $a$, depending on the expectation values $x_{a}$ of the complex scalars in the vector multiplet of the 3d gauge theory on a circle.
The hand-saw structure of the quiver implies we can write the superpotential as\linebreak
\beq\label{dW}
\partial_x {\cal W}_a(x) = p_a(x)-p_{a-1}(x)
\eeq
where $p_a(x)$ to contains the contributions of the fundamental chirals at node $a$: this reflects the fact that the contribution of the anti fundamental chirals at the node $a$ and of the fundamental chirals at the previous node, coincide. Furthermore,

\beq\label{3dpV}
e^{p_{a}(x)} = V_a(e^x) =  {e^{\zeta_a }\over \prod_{i=1}^{\ell} ( 1- e^x/  f_i^{(a)}) },
\eeq
where we denoted

$$
f_i^{(a)} = e^{x_{*,i}^{(a)}}, \qquad i=1, \ldots \ell
$$
Note that each function $p_a(x)$ is determined solely by the corresponding $U(\ell)$ subgroup of $F_{3d}$. The $n+1$ functions $p_a(x)$ with $\ell$ poles each are compactly in encoded in the curve

$$
S\;\;:\qquad\qquad \prod_{a=0}^{\ell}(e^p - e^{{p_a}(x)})=0
$$
The curve is exactly the same as the $S$-curve of the 5d theory in the previous section, provided we identify parameters appropriately.
The positions of the $\ell$ punctures $x^{(a)}_{i, *}$, $i=1, \ldots \ell$ on the $a$-th sheet are the background vector multiplet scalars in both the 5d and the 3d theory interpretation of the $S$-curve.\footnote{The scalars in vector multiplets of both the 5d ${\cal N}=1$ and 3d ${\cal N}=2$ supersymmetry are real, and get complexified by compactifying the theories on a circle, as we do here.
The background vector multiplets gauge the flavor symmetry associated to the fundamental matter multiplets; these are the hypermultiplets in the 5d case and chiral multiplets in 3d.} Each of the $n+1$ $U({\ell})$ subgroups of  $F_{3d}$ corresponds to one of the sheets of the $S$-curve in \eqref{5dcurve2}. The parameters $\zeta_a$ in \eqref{5dcurve2} determine the Fayet-Illiopolous parameters in 3d, and gauge couplings in 5d. They enter the twisted superpotential linearly, and hence affect $p$ by a constant shift.


%

\subsection{${\cal V}_{3d}$ from String Theory}

We will now explain how the 3d theory ${\cal V}_{3d}$ arises in string theory.  Later on, we will use this to show that it describes dynamics of vortices in ${\cal T}_{5d}$. It turns out that ${\cal V}_{3d}$ arises on D3 branes in IIB string theory compactified on a blowup of $Y_S$ in \eqref{5dcurveCY}

$$
Y_S\;\;:\qquad\qquad \prod_{a=0}^{\ell}(e^p - e^{{p_a}(x)})=uv,
$$
the Calabi-Yau based on the $S$-curve -- the same Calabi-Yau which we first saw in the previous section. One can view $Y_S$ as a family of $A_n$ surfaces, one for each point in the $x$-cylinder. After blowing up, at each $x$, there are $n$ $S^2$'s of non-zero area. The areas of the $S^2$'s vary with $x$: they are minimized, yet non-zero, where the sheets of the $S$-curve intersect. The theory we called ${\cal V}_{3d}$ arises by wrapping $N_{(a)}$ D3 branes on the $a$-th $S^2$ class in the $A_n$ surface. Let us explain why this is the case.

\subsubsection{The ${\ell}=0$ Case}

Consider first the case when  ${\ell}=0$. In this case the Calabi-Yau is a direct product of the $A_n$ singularity

$$
\;\;\qquad\qquad \prod_{a=0}^{\ell}(e^p - e^{{\zeta_a}})=uv.
$$
and $C$, where $C$ is a cylinder parameterized by $x$. At the same time, ${\cal V}_{3d}$ is a 3d $A_n$ quiver theory with ${\cal N}=4$ supersymmetry, compactified on $S^1$. The quiver of ${\cal V}_{3d}$, in this case, is the quiver discovered in \cite{Douglas:1996sw}, describing fractional branes wrapping the $S^2$'s in an $A_n$ surface. We take the D3 branes to be transverse to $C$ -- so the compact scalar in their world volume describes the position on $C$.\footnote{At this level, we could have chosen the branes to wrap the $S^1$ equally well, but consideration of general $\ell$ will show the perspective we chose is the right one.} The rank $N_a$ of the quiver gauge group associated to $a$-th node of the quiver is the number of the D3 branes wrapping the corresponding $S^2$ generating the second homology group of the $A_n$ surface.

The one subtlety is that the Dynkin diagram that is relevant is the Dynkin of the ordinary $A_n$ Lie algebra, rather than the affine one. Affine quiver arises only at the orbifold point -- and for $\zeta_a$ constant, but otherwise generic, one is away from the orbifold point. Away from the orbifold point  D-branes wrapping the affine node are not mutually supersymmetric with the rest, so one sets the rank of the gauge group on the affine node to be zero \cite{Douglas:1996sw}.

\subsubsection{The ${\ell}\neq 0$ Case}

For ${\ell} \neq 0$, ${\cal V}_{3d}$  has only ${\cal N}=2$ supersymmetry. Integrating out the charged fundamental matter, the theory gets a non-trivial effective twisted superpotential for the scalar in  the vector multiplet. The geometric interpretation of this was given in \cite{Cachazo:2001gh}, following \cite{MV}. The non-trivial superpotential means that the $A_n$ singularity is fibered non-trivially over $C$ with its complex coordinate $x$ identified with scalar in the vector multiplet, complexified by the holonomy. More precisely, the superpotential ${\cal W}_{(a)}(x)$ for the  $a$-th node means that \cite{Cachazo:2001gh}

$$
{\partial_x}{\cal W}_{(a)} = {1\over 2\pi i} \int_{S^2_{(a),x}} \omega^{2,0} .
$$
Here,  $\omega^{2,0}$ is the holomorphic $(2,0)$ form on the $A_n$ surface, taken in the fiber over the point $x$ on $C$, and
$S^2_{(a),x}  $ is the $S^2$ corresponding to the $a$-th node, sitting above a point $x$ on $C$. The integral on the right measures the symplectic volume of the $S^2_{(a)}$; the curve can be holomorphic only where this vanishes. When this happens, and the $S^2_{(a)}$ is holomorphic, the theory has a supersymmetric vacuum. These in turn are  the minima of the superpotential  ${\cal W}_{(a)}$. This is the content of the above formula, discovered first in \cite{Witten:1997ep}.

In the present case, taking the effective twisted superpotential to be of the form  given in \eqref{dW}

$${\partial_x}{\cal W}_{(a)}=p_{a}(x)-p_{a-1}(x),$$
we see that it exactly corresponds to the D3 branes wrapping the $S^2$'s in the blowup of $Y_S$, as claimed.

From this, we can deduce the M-theory realization of ${\cal V}_{3d}$ using string dualities which relate IIB string on $Y_S$ to M-theory with M5 brane wrapping the $S$-curve. Blowing up singularities of $Y_S$, corresponds, in M5 brane language to separating in the 3 transverse directions (these are the $x^7,x^8, x^9$ directions in the notation of \cite{W7}). Moreover, the D3 branes wrapping the ${\mathbb P}^1$'s map to M2 branes stretching between the consecutive pairs of M5 branes.

\subsection{${\cal V}_{3d}$ Lives on Vortices in ${\cal T}_{5d}$ }

The 3d ${\cal N}=2$ gauge theory ${\cal V}_{3d}$ is a theory on charge $N_{a}$ vortices in on ${\cal T}_{5d}$. This can be seen as follows.

The vortices in question are non-abelian generalization of Nielsen-Olson vortex solutions. BPS vortex solutions arise on (baryonic) Higgs branches of unitary gauge theories such as ${\cal T}_{5d}$. Examples of these were constructed in \cite{HananyTong, HananyTong2}. The BPS tension is set by the value of the FI parameters. The net BPS charge of the vortex is

$$N_{(a)}=-\int {\rm Tr} F_{(a)},
$$
where $F_{(a)}$ is the field strength of the corresponding gauge group factor and the integral is taken in the 2 directions transverse to the vortex.\footnote{Usually, the gauge theories on M5 branes wrapping Riemann surfaces are said to be of special unitary type, rather than unitary type. There is no contradiction; the $U(1)$ centers of the gauge groups that arise on branes are typically massive by Green-Schwarz mechanism. This does not affect the BPS tension of the solutions, see for e.g. discussion in \cite{Douglas:1996sw}.}

From the perspective of IIB, going on the Higgs branch of ${\cal T}_{5d}$ \cite{Strominger:1995cz,Greene:1995h} corresponds to blowing up singularities of $Y_S$. The vortices are D3 branes wrapping the $n$ $S^2$ classes on the blowup \cite{Greene:1996dh, Hori:1997zj}.  The BPS tension of the wrapped brane matches the BPS tension of the vortex, and so does its charge.   From the perspective of $n+1$ M5 branes wrapping $C$, the vortices are M2 branes stretching between pairs selected out of $n+1$ M5 branes.\footnote{One should not confuse the vortices here with surface operators in the gauge theory, studied for example in \cite{Alday:2009fs,Dimofte:2010tz}. The surface operators are solutions on the Coulomb branch, with infinite tension. From the M5 brane perspective, surface operators are semi-infinite M2 branes ending on M5's, while vortices we study are M2 branes which end on pairs of M5's and are finite extent, in one of their directions.}  This is exactly how ${\cal V}_{3d}$ arises in string theory.

\subsubsection{The Hand-Saw Quiver}

The hand-saw quiver of ${\cal V}_{3d}$ given in figure \ref{HandsawFigure} was recently studied by Nakajima in \cite{NHS}. We just gave a physical explanation for how this quiver arises in string theory. \footnote{For another embedding, see \cite{YK}, in a somewhat different context.} This is a generalization of the work of \cite{Douglas:1996sw} which applies for the ${\cal N}=4$ theory at $\ell=0$, to arbitrary $\ell$. The upshot is that this more general quiver corresponds to fractional D3 branes on $A_n$ singularity fibered over ${C}$ as opposed to a direct product ($C$ could be a complex plane, or a cylinder in our case, depending on whether the quiver theory is or is not compactified on a circle).

In \cite{NHS}, the Higgs branch of ${\cal V}_{3d}$ is called the {\it handsaw quiver variety} ${\cal D}_{{\cal V}_{3d}}$.
The fact that ${\cal V}_{3d}$ describes vortices in ${\cal T}_{5d}$ implies that ${\cal D}_{{\cal V}_{3d}}$ is the moduli space of charge $N_a$ vortices in ${\cal T}_{5d}$.  This statement was in fact proven in \cite{NHS} (although the original result is older, going to late 80's).
Briefly, \cite{NHS} proves that is ${\cal D}_{{\cal V}_{3d}}$ is isomorphic to the so called {\it parabolic Laumon space} associated to a flag variety $X=Fl(d_1,\ldots, d_n; d_{n+1})$, where $d_a =a \ell$. The parabolic Laumon space is roughly the moduli space of holomorphic maps from ${\mathbb P}^1$ to $X$ of degrees $(N_1, \ldots, N_n)$. One requires that the point at infinity of the ${\mathbb P}^1$ gets mapped to a specific point of the of flag variety $X$.

It is easy to see that this produces vortex solutions of ${\cal T}_{5d}$. To produce a solution of vortex equations in this theory one needs \cite{Wittenphases} a holomorphic vector bundle with gauge group $G_{5d}=\prod_{a=1}^{n}U(d_a)$ and first Chern class $\int {\rm Tr} F_a = -N_a$, specifying the vortex charge, together with a set of matter fields of the quiver of ${\cal T}_{5d}$ which are holomorphic sections of the bundle, solving the F-flatness conditions, with fixed values at infinity. A way to solve the F-flatness conditions is to set all the matter fields of the 5d quiver to zero identically, except for those transforming in representation $(d_1, {\overline d_2})\oplus \cdots \oplus (d_n,{\overline d_{n+1}})$, corresponding to the horizontal arrows of the quiver in Fig.\ref{GaugeQuiver} (we only do the all full punctures case here). The space of remaining bifundamental matter fields modulo the complexified gauge group is, per definition \cite{Donagi:2007hi}, the flag variety $X$ appearing in Nakajima's work. Matter fields provide holomorphic maps from ${\mathbb C}$ -- a ${\mathbb P}^1$ with marked point -- to $X$, with degrees $(N_1, \ldots , N_n)$.
The space of such solutions is clearly the parabolic Laumon space of $X$.

\subsection{Partition Function of ${\cal V}_{3d}$}
Consider ${\cal V}_{3d}$ in $\Omega$-background -- this is the three manifold $M_q$

$$
M_q = ({\mathbb C} \times S^1)_q.
$$
In $M_q$, as we go around the $S^1$, we simultaneously rotate the complex plane by $q$, and turn on the flavor symmetry group in \eqref{3df},
as well as the $U(1)_R$-symmetry twist, to preserve supersymmetry. If we denote the element we twist by ${\bf g}$, the partition function of the theory on $M_q$ computes the index

\beq\label{3dtrace}
{\cal Z}_{{\cal V}_{3d}}={\rm Tr} (-1)^F {\bf g}.
\eeq
${\bf g}$ is a combination of $SO(2)$ rotations of ${\mathbb C}$  and the holonomy of the $U(1)_R$ symmetry by $q$, and holonomy for the global symmetry including $U(1)_t$.

The partition function ${\cal Z}_{{\cal V}_{3d}}$ can be computed as the integral over the Coulomb branch:

\beq\label{3dpart}
{\cal Z}_{{\cal V}_{3d}}(S; N)=\int dx \; I_{{\cal V}_{3d}}(x;S).
\eeq
The integrand $I_{{\cal V}_{3d}}(x)$ is the value of the index \eqref{3dtrace} if we view the Coulomb branch moduli $x$ as fixed -- this renders the theory free. The integration over the Coulomb branch moduli, denoted by $"\int dx\;"$
restores $G_{3d}$ as a gauge, instead of the global symmetry. The integral runs over the Coulomb branch moduli for each of the $n$ gauge group factors in \eqref{3dg}:

\beq
"\int dx\;" =
{1\over |W_G|}\prod_{a=1}^n \int d^{N_a}x^{(a)}.
\eeq
The integrand receives contributions from massive matter:

\beq
 I_{{\cal V}_{3d}}(x;S)=
 \;\; \prod_{a=1}^n \Bigl(\prod_{i=1}^{\ell} \Phi_{H^{a}_i}(x^{(a)})\Bigr)  \;
 \Phi_{V_a}(x^{(a)})\,
 \Bigl(\prod_{b>a} \Phi_{H^{a, b}}(x^{(a)}, x^{(b)}) \Bigr)\,  e^{ \zeta_a \,{\rm Tr}\, x^{(a)}/\hbar}
\eeq
There is a universal contribution to the index, for each $U(N_a)$ gauge group, coming from the vector multiplet of the would-be ${\cal N}=4$ supersymmetry:

\beq
\Phi_{V_a}(x^{(a)}) = \prod_{1\leq I \neq J \leq N_{(a)}} { \varphi(  e^{x^{(a)}_I-x^{(a)}_J})\over \varphi(   t \,e^{x^{(a)}_I-x^{(a)}_J})},
\eeq
The numerator is the W-boson contribution, the denominator comes from the adjoint chiral of mass $\epsilon$, $t=e^{\epsilon}$.
The bifundamental hypermultiplet, corresponding to a a pair of nodes $a$ and $b$, with a link between them gives:

\beq\label{basic}
\Phi_{H^{a,b}}(x^{(a)},x^{(b)})= \prod_{1\leq I \leq N_{(a)}}   \prod_{1\leq J\leq N_{(b)}} \Bigl(  {\varphi(  t \, e^{x^{(a)}_I-x^{(b)}_J})\over \varphi( \; e^{x^{(a)}_I-x^{(b)}_J})}\Bigr)^{I_{ab}},
\eeq
To indicate the fact that only the pairs of nodes which are linked contribute, in the partition function the contribution of $\Phi_{H^{a,b}}$ gets raised to the  power of the incidence matrix $I_{ab}$, where $I_{ab} = C_{ab}+ 2 \delta_{ac}$ and $C_{ab}$ is the Cartan matrix. In addition, at each node there are ${\ell}$ chiral multiplets in fundamental representation, and $\ell$ in anti-fundamental representation. A flavor $H_a$ in fundamental representation of $U(N_a)$ gauge group contributes

\beq\label{basic}
\Phi_{H_a}(x^{(a)}) =  \prod_{1\leq i \leq \ell }\;\;\prod_{1\leq I \leq N_a} {\varphi( v \, e^{x^{(a)}_I}/f_i^{(a)})\over \varphi( v^{-1} \;
e^{x^{(a)}_I}/f^{(a+1)}_i)}.
\eeq
In all of the above, the $t$ and the $v=(q/t)^{1/2}$ factors are related to the charges under the $U(1)_R$, $U(1)_t$ and Lorentz symmetries, up to redefinitions of the $x$'s and the masses. We defined

$$
\varphi(z) = \prod_{n=0}^{\infty}(1-q^n z).
$$

The mass parameters $f^{(a)}_i$ are determined from the $S$-curve. Namely, on the one hand, the $S$-curve encodes the effective twisted superpotential in the limit in which $\hbar=\epsilon_1$ and $\epsilon=\epsilon_2$ go to zero, where $q=e^{R\hbar}, t=e^{-R\epsilon}$, given by ${\cal W}(x)$ in \eqref{3dW}. On the other hand, the integrand is determined by ${\cal W}$:

\beq
 I_{{\cal V}_{3d}}(x;S )\stackrel{{\hbar}, {\epsilon} \rightarrow 0}{\longrightarrow}\;\exp\left({1\over \hbar}{\cal W}(x; S)\right),
\eeq
which is a consequence of the fact that $\varphi(z)\stackrel{{\hbar}, \epsilon \rightarrow 0}{\longrightarrow} \exp({1\over \hbar}Li_2(z))$.

The partition function \eqref{3dpart} has the form of a matrix integral, in eigenvalue form, with measure $\Phi_{V_a} $ for the $N_{(a)}$ eigenvalues $x^{(a)}$. The matrix integrals are of refined Chern-Simons type \cite{AS, AS2}. This is natural given that their string theory origin is the same.

\subsubsection{Dependence On The Contour}

To fully specify the partition function ${\cal Z}_{{\cal V}_{3d}}$, we need to specify the contours of integration in \eqref{3dpart}. The choice of contours is crucial: in the next section, when we make contact with Toda CFT, the dependence of ${\cal Z}_{{\cal V}_{3d}}$ on the contours will lead to extra parameters for the conformal vertex operators.

The dependence on the contour reflects the choice of the vacuum of the 3d gauge theory at infinity of $M_{q}$\footnote{One may be able to further generalize this by the dependence on the twistorial parameter, see the recent work \cite{tt1, tt2}.}. Classically, the vacua are located at the intersections of the different components of the $S$-curve:

\beq\label{3dc}
V_a(e^x)=V_b(e^x).
\eeq
The approximate locations of the solutions to \eqref{3dc} can be read off the web diagrams which capture the 3d limits of the curves. This also allows to count the vacua: for every pair $a,b$ the curves intersect over $\ell$ points\footnote{Assuming all punctures are full; otherwise the count is analogous but different.} in $x$, hence, there are $\ell n(n+1)/2$ vacua all together.

The choice of contours ends up effectively breaking the gauge symmetry,

\beq\label{break}
G_{3d} = \prod_{a=1}^n U(N_{(a)}) \quad \longrightarrow \quad \prod_{a=1}^n \prod_{i=1}^{d_{a}} U(N_{a,i})
\eeq
where we labeled the contours by pairs $a, i$ and associated a subset $x^{(a), i}$ consisting of $N_{a,i}$ integration variables to each. The symmetry breaking requires $\sum_{i=1}^{d_{a}}N_{a, i} = N_a$ and $d_a = \ell (n-a+1)$.

Each contour is a closed loop in the $z=e^{-x}$ plane, encircling $z=0$ and positions of poles of the integrand. Without loss of generality, the partition function can be computed by integrating out the groups of variables one by one, starting from $x^{(1)}$, finishing by $x^{(n)}$. For the last group $x^{(n)}$, when all other variables are already integrated out and hence all singularities associated with $\Phi_{H^{a,b}}$ are smeared, the only non-homotopic contours are those going around the poles of $\Phi_{H_n}$,
$$
\underbrace{ \Big[ 0, \big(v f^{(n+1)}_{i}\big)^{-1} \Big]}_{N_{n,i}}, \ \ \ i = 1, \ldots, \ell
$$
For the second to last group $x^{(n-1)}$, one has two choices: $x^{(n)}$ is not yet integrated out, hence both $\Phi_{H^{n-1,n}}$ and $\Phi_{H_n}$ contribute to the pole structure. The contours associated to these are, respectively,
$$
\underbrace{\big[ 0, x^{(n), i} \big]}_{N_{n-1,2i-1}} \ {\rm and} \ \underbrace{\Big[ 0, \big(v f^{(n)}_{i}\big)^{-1} \Big]}_{N_{n-1,2i}}, \ \ \ i = 1, \ldots, \ell
$$
Similarly, for the third to last group $x^{(n-2)}$, one has three choices,
$$
\underbrace{\big[ 0, x^{(n-1), 2i-1} \big]}_{N_{n-2,3i-2}} \ {\rm and} \ \underbrace{\big[ 0, x^{(n-1), 2i} \big]}_{N_{n-2,3i-1}} \ {\rm and} \ \underbrace{\Big[ 0, \big(v f^{(n-1)}_{i}\big)^{-1} \Big]}_{N_{n-2,3i}}, \ \ \ i = 1, \ldots, \ell
$$
and so forth. At each step, one more choice appears, due to the poles of the bifundamental contributions $\Phi_{H^{a,a+1}}$. This explains why, with $n$, the number of contours grows as a triangular number $\ell n(n+1)/2$.

In the language of IIB on $Y_S$, solutions to \eqref{3dc} is where the minimal $S^2$'s are located. In M-theory language, this is where the M2 brane vortices have minimal, BPS mass.  Deciding how many branes is located at each solution of \eqref{3dc} is reflected in the breaking pattern in \eqref{break}.

\section{$A_n$ Toda CFT and Triality}

In this section, we review aspects of the 2d $A_n$ Toda CFT and the Coulomb gas, or Dotsenko-Fateev (DF), formulation for conformal blocks of the theory. The theory has a $W_{n+1}$-algebra symmetry.\footnote{See
\cite{Fateev:2007ab} for a recent review, and \cite{Fateev:1987zh} for the original paper constructing $W$-algebra of $A_n$ type and a theory realizing the symmetry.} The $W$-algebra admits a $q$-deformation, and this results in deformation of the conformal blocks of the $A_n$ Toda theory. In the $n=1$ case, the Toda CFT reduces to Liouville field theory, and the $W_2$ algebra is just the Virasoro algebra.

We will show that $q$-deformed conformal blocks of $A_n$ Toda CFT on a sphere, with primary operators inserted, agree {\it manifestly} with the partition function of the vortex theory ${\cal V}_{3d}$. This will prove part $ i.$ of $A_n$-triality.

\subsection{Review of $A_n$ Toda and its W-algebra symmetry.}

The $A_n$ Toda field theory can be written in terms of $n+1$ free bosons $\phi_a$ in two dimensions, with canonical kinetic term, a background charge contribution and the Toda potential that couples them. The action of the theory is
$$ S_{Toda} = \int dz d{\bar z} \;\sqrt g \; [g^{z {\bar z}}(\partial_z \phi,\partial_{\bar z} \phi) +  Q (\rho, \phi)R + \sum_a e^{(\phi, e_{(a)})} ].
$$
The inner product is the standard inner product on ${\mathbb R}^{n+1}$, $\rho$ is the Weyl vector of the $A_n$ Lie algebra so $(\rho, \phi) = {1\over 2} \sum_{a=1}^{n+1} (n-2a+2) \phi_{a}$. The vectors $e_{(a)}$ stand for the $n$ simple roots,

$$(\phi, e_{(a)}) = \phi_a-\phi_{a+1}.
$$
The Toda potential couples $n$ of the bosons, but since one boson $\phi_0= \sum_{\alpha} \phi_{\alpha}$ remains decoupled from the rest,  the correlation functions will factorize. In defining the Toda field theory one usually decouples this degree of freedom from the outset; we will not do that.

\begin{figure}[h!]
\begin{center}
\includegraphics[width=0.7\textwidth]{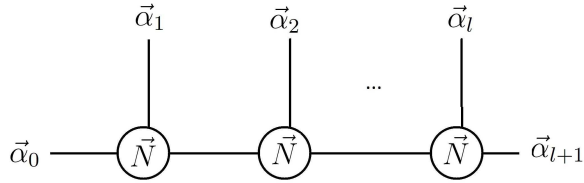}
\end{center}
\caption{$(l+2)$-point spherical comb conformal block of $A_n$ Toda theory. }
\label{CFTversusQFTb}
\end{figure}

The primary vertex operators of $A_n$ Toda are of the form

$$
V_{\alpha}(z) = e^{(\alpha, \phi(z)/b)}
$$
where $\alpha =  (\alpha_{1},\alpha_{2},\ldots,\alpha_{n+1})$, for $\alpha_{a}$ generic complex numbers. Conformal blocks on a sphere with $k$ punctures, in the channel corresponding to the Fig.\ref{CFTversusQFTb} are given, implicitly, by the following free field correlator,

\beq\label{expect}
\left< \ V_{\alpha_1}(z_1) \ldots V_{\alpha_k}(z_k) \;\;\prod_{a=1}^{n} Q_{(a)}^{N_a} \ \right>
\eeq
where the screening charges

$$
Q_{(a)} = \oint dx \ S_{(a)}(x)
$$
are the integrals over the screening currents $S_a(x)$, one for each simple root,

$$
S_{(a)}(z) = e^{b (e_{(a)}, \phi(z))}.
$$
To derive \eqref{expect} one treats the Toda potential as a perturbation. Bringing down powers of the Toda potential (and picking out a chiral half of the correlator), inserts screening charge integrals. One finds that
\eqref{expect} vanishes unless a "charge conservation constraint" is satisfied

$$
2Q -  b \sum_{a=1}^n N_a e_{(a)}=\sum_{i=1}^k \alpha_i
$$
correlating the net charge of the vertex operator insertions with the number of screening charge integrals $N_a$. This constraint can be found directly from the path integral, by integrating over the zero modes of the bosons \cite{Fateev:2007ab}. We will place a vertex operator at infinity of the $x$ plane, and then the equation determining the momentum of the operator at infinity, in terms of the momenta of the $\ell+1$ remaining vertex operators at finite points and numbers of screening charge integrals.

One can easily compute the expectation value of the correlator in \eqref{expect} by using the free field expansions of the bosons

$$
\phi^{a}(z)= \phi^{a}_0+h_0^{a}\,\mbox{log}\,z+\sum_{k\neq 0}h_k^{a}\frac{z^{-k}}{k}.
$$
where
$\psi^{a}_0$ is a constant, and
$h^{a}_m$ , for $a$ running from $1$ to $n+1$, generate $n+1$ commuting copies of the standard Heisenberg algebra
relations

\beq
\label{heisenberg}
[h^{a}_{k}, h^{b}_{m}] = {k} \,\delta_{k+m,0} \,\delta^{a,b}, \qquad a,b=1,\ldots, n+1,
\eeq
where $k, m\in {\mathbb Z} $.
%
%
%
%

The Toda CFT has a $W_{n+1}$ algebra symmetry. The $W_{n+1}$ algebra is an algebra generated by currents of spin up to $n+1$ built out of the free fields and it includes the Virasoro algebra as a subgroup (the $W_2$ algebra is the Virasoro algebra itself).
%
%
The $W$-algebra generators commute with the integrals of the screening charge operators, expressing the fact that the symmetry of the free boson CFT is preserved by adding the Toda potential.

From Toda CFT perspective, the $S$-curve \eqref{4dcurve2} that captured the data of ${\cal T}_{4d}$ arises very naturally -- it encodes the insertions of vertex operators in the free CFT, before we bring down the screening charge integrals.

\subsubsection{$q$-deformation of $A_n$ Toda}

The $q-$deformed $W$-algebra generators are defined by the property that they commute with the integrals of $q$-deformed screening charges $S_a$ \cite{FR1}

$$
\begin{aligned}
S_{(a)}(z) = \ :\exp\left(\sum_{k\neq 0}\frac{\alpha_k^a\  z^{-k}}{q^{k/2}-q^{-k/2}}\right):
\end{aligned}
$$
where $\alpha_k^a$'s are simple linear combinations of the boson modes $h_k^a$,
$$
\begin{cases}
\alpha_k^a &=\left(t^{-k/2}-t^{k/2}\right)\left(v^{-k}h_k^a-h_k^{a+1}\right)/k\qquad\ \ \ \ (k>0) \\
\alpha_{-k}^a &=\left(q^{-k/2}-q^{k/2}\right)\left(h_{-k}^a-v^{k}h_{-k}^{a+1}\right)/k\qquad\ \ \ \, (k>0)
\end{cases}
$$
their explicit definition being the same as in \cite{FR1}. In particular, the $h_k^a$ modes in what follows still satisfy the Heisenberg algebra \eqref{heisenberg}.
%
%
%
%
We therefore obtain the following form for the screening operators:
$$
\begin{aligned}
S_{(a)}(z) &= \ :\exp\Bigl(-\sum_{k>0} {1-t^k\over 1-q^k}h^{a}_k\; \frac{z^{-k}}{k} +\sum_{k>0}   \;h^{a}_{-k} \;\frac{z^{k}}{k}\Bigr):\\
&\times  : \exp\Bigl(\sum_{k>0}{1-t^{k}\over 1-q^{k}}  v^k\,h^{a+1}_k \; \frac{z^{-k}}{k} -\sum_{k>0}   \;v^{k}\,h^{a+1}_{-k}\;\frac{z^{k}}{k}\Bigr):\ \
\end{aligned}
$$
Note that $S_{(a)}(z)$  contains both the $(a)$-th and $(a+1)$-th sets of modes. In the limit where $q$, $t$ go to $1$, with $t=q^{\beta}$ and $\beta$ fixed, this reduces, after a rescaling of $h$'s to the ordinary $W$ algebra. (Note that $\beta=-b^2 =- \epsilon_1/\epsilon_2$.)

The $q$-deformed primary vertex operators can be obtained from
$$
{ V}_{\alpha}(z)= :\exp\Bigl(\sum_{k>0}\sum_{a=1}^{n+1}  {1\over 1-q^k}  \;h^{a}_{k} \;q^{k\alpha_a} \;\frac{z^{-k}}{k}+\sum_{k>0}\sum_{a=1}^{n+1} {q^k\over 1-t^k}\;h^{a}_{-k}\; q^{-k\alpha_a}\;\frac{z^{k}}{k} \Bigr):
$$
As written, the operator does not have a good $q\rightarrow 1$ limit -- only its two point functions with screening charges do. There is an easy fix. The properly defined vertex operator is obtained from ${V}_{\alpha}$ by removing central terms -- terms that commute with the screening operators; what we remove is uniquely determined by asking that the resulting operator has a well defined $q\rightarrow 1$ limit. We will use the same notation $V_{\alpha}$ for both operators, for simplicity. In fact, after a specialization of our operator to specific momenta, our answer agrees with the vertex operators  defined in \cite{FR1}\footnote{Operators ${V_{i}}^{RF}$ of \cite{FR1} are defined for all $i$ from $1$ to $n+1$, and correspond to the momenta $\alpha=(1,\ldots,1,0,\ldots,0)$, where the first $i$ entries are $1$, and the rest are $0$.}.

The $q$-deformation deforms to operators from those of the Toda CFT, by $q$ and $t$ dependent factors. From the algebra of the operators, it is easy to work out the $q$-deformation of the conformal block of the $A_n$ theory:

\beq\label{expect5d}
{\cal B}_{{\rm q-Toda}}(S; N) = \left< V_{\alpha_1}(z_1) \ldots V_{\alpha_k}(z_k) \;\;\prod_{a=1}^{n} Q_{(a)}^{N_{a}} \right>
\eeq
Let us denote by $\int dx$ collectively the integrals over the positions of the screening charges,
$$
" \int dx " =\prod_{a=1}^{n} {1\over N_a!} \oint \prod_{I=1}^{N_{a}} dx^{(a)}_{I}
$$
so that

$$
{\cal B}_{{\rm q-Toda}}(S) =\int dx\; I_{{\rm q-Toda}}(x),
$$
where the integrand
$$
I_{{\rm q-Toda}}(x) = \left< \ V_{\alpha_1}(z_1) \ldots V_{\alpha_k}(z_k) \;\;\prod_{a=1}^{n} \prod_{I=1}^{N_{a}} S_{(a)}\big(x^{(a)}_{I}\big) \ \right>
$$
is the value of the correlator with screening charge contributions.

\subsection{$A_n$-Triality Part $i.$}

The claim in the introduction is that the Coulomb gas expression for the conformal block is the same as the partition function of the 3d gauge theory ${\cal V}_{3d}$, in $\Omega$-background. The partition function of ${\cal V}_{3d}$ is given in \eqref{3dpart},

$$
{\cal Z}_{{\cal V}_{3d}}(S; N)=\int dx \; I_{{\cal V}_{3d}}(x;S).
$$
We will now prove that

$$
 I_{{\cal V}_{3d}}(x)=I_{{\rm q-Toda}}(x).
$$
We will show that the numbers of screening charge integrals map to ranks of the 3d gauge groups; inserting a vertex operator in $q$-Toda maps to coupling the 3d gauge theory ${\cal V}_{3d}$ to an additional flavor, for each node of the 3d quiver.

The 3d partition function is a product of terms
%
associated to vector multiplets and bifundamentals -- these are contributions of $\Phi_{V_a}$ and $\Phi_{H^{a,b}}$ -- and the fundamental flavors $\Phi_{H^a}$ and the FI terms (the exponential factor). The Toda integrand is also a product of terms from two-point functions

$$
I_{{\rm q-Toda}}(x) = r \times \prod_{a=1}^n  \prod_{i=1}^k I_{{a}}\big(x^{(a)};z_i, \alpha_i\big)\times \prod_{1\leq a<b \leq n} I_{a, b}\big(x^{(a)}, x^{(b)}\big)
$$
where

$$I_{a, b}\big(x^{(a)}, x^{(b)}\big) = \prod_{I=1}^{N_{a}}\prod_{J=1}^{N_{b}} \left< \ S_{(a)}\big(x^{(a)}_{I}\big) \; S_{(b)}\big(x^{(b)}_{J}\big) \ \right>
$$
and
$$I_{{a}}\big(x^{(a)};z, \alpha\big) = \prod_{I=1}^{N_{a}} \left< \ {V}_{\alpha}(z) \; S_{(a)}\big(x^{(a)}_{I}\big) \ \right>
$$
The normalization factor $r$ includes  the contributions from the correlators $\langle  {V}_{\alpha_i}(z) \; {V}_{\alpha_j}(w) \rangle$.

The contributions of vector and bifundamental hypermultiplets to $I_{{\cal V}_{3d}}$ map to contributions of two point functions between screening charges to $I_{{\rm q-Toda}}(x)$, and contributions of matter fields map to two-point functions of screening charges and vertex operators. The only non-trivial two-point functions that enter $I_{{\rm q-Toda}}$ are

\begin{align*}
\left< \ S_{(a)}\big(x^{(a)}_{I}\big) \; S_{(a)}\big(x^{(a)}_{I}\big) \ \right> &=  {\varphi\big( x^{(a)}_{J} / x^{(a)}_{I} \big) \over\varphi\big( t x^{(a)}_{J} / x^{(a)}_{I} \big) }{\varphi\big( x^{(a)}_{I} / x^{(a)}_{J} \big) \over\varphi\big( t x^{(a)}_{I} / x^{(a)}_{J} \big)},\\
\left< \ S_{(a)} \big(x^{(a)}_{I}\big)\; S_{{(a+1)}}\big(x^{(a+1)}_{J}\big) \ \right> &= {\varphi\big( u \, x^{(a+1)}_{J} / x^{(a)}_{I} \big) \over\varphi\big( v \, x^{(a+1)}_{J} / x^{(a)}_{I} \big) },\\
\left< \ {V}_{\alpha}(z)\;S_{(a)}(x_{(a),I}) \ \right> &={{\varphi( q^{1-\alpha_{a}} z/x^{(a)}_{I} )}\over\varphi(v^{-1}\, q^{1-\alpha_{a+1}}\,z/x^{(a)}_{I})},
\end{align*}
where $u=(qt)^{1/2}$. In particular $I_{a,b} \big(x^{(a)},x^{(b)}\big) = 1$ for $b\neq a, a+1$. Therefore
$$
\Phi_{V_{a}} \big(x^{(a)}\big) = I_{a, a}\big(x^{(a)}, x^{(a)}\big),
$$
$$
\Phi_{H^{a,b}} \big(x^{(a)}, x^{(a)}\big) = I_{a, b}\big(x^{(a)}, x^{(b)}\big), \qquad a\neq b
$$
$$
\Phi_{H^{a}_i} \big(x^{(a)}\big)=I_{{a}}\big(x^{(a)};z\big)
$$
under the following map of parameters:
{\renewcommand{\arraystretch}{1.5}
\renewcommand{\tabcolsep}{0.2cm}
\begin{center}
\begin{tabular}{r|l}
  \hline
  q-Toda\; & \;3d gauge theory \\ \cline{1-2}
  $x^{(a)}_I\;\;\;\;$ &  $\;\;\;e^{-x^{(a)}_I}\;v^{-a}$ \\
  $q^{\alpha_a-1}\;z_i^{-1}\;$ & $\;\;\;\;f_i^{(a)}\;v^{a-1}$ \\
\end{tabular}
\end{center}}

\section{Gauge/Vortex Duality and $A_n$-Triality}

Gauge/vortex duality relates a 4d ${\cal N}=2$ gauge theory in a variant of 2d ${\Omega}$-background, and the 2d ${\cal N}=(2,2)$ theory on its vortices. The duality relates ${\cal T}_{5d}$ and  ${\cal V}_{3d}$, compacted on a circle. The extra circle naturally enters in defining the ${\Omega}$-background. A consequence of the gauge/vortex duality is the equality of the partition functions of the two theories ${\cal Z}_{{\cal T}_{5d}}$ and ${\cal Z}_{{\cal V}_{3d}}$. We'll first review the gauge/vortex duality and then prove that it indeed relates the partition functions.\footnote{The fact that  the BPS spectra of the two theories are related was observed in \cite{Dorey, DHT, HananyTong, HananyTong2, Shifman:2004dr}. The duality was first proposed in \cite{DH1, DH2} based on the fact that, upon turning on $\Omega$ background and specific values of Coulomb branch moduli, the super potentials of the 2d and the 4d theory agree. The physical explanation for gauge/vortex duality we just gave appeared first in \cite{SimonsTalk}.}

\subsection{Review of Gauge/Vortex Duality}

On one side of the duality is a 4d ${\cal N}=2$ theory, compactified on a {\it two}-dimensional ${\Omega}$-background; this system is studied in \cite{NS2}.  Relative to section two, we set $\epsilon_1=\hbar=0$ momentarily (since the duality holds for any $\hbar$), and $\epsilon_2=\epsilon$. The ${\Omega}$-background is an alternative to compactification \cite{NS} -- so this  results in a 2d theory ${\cal N}=(2,2)$ theory with infinitely many massive modes, with masses spaced in multiples of $\epsilon$. In addition, we turn on vortex charge is $\int_D F_i =-N_i$ where $i$ labels a $U(1)$ gauge field in the IR, and $F_i$ is the corresponding field strength. Here, $D$ is the cigar, the part of the 4d space time with 2d $\Omega$ deformation on it. Without the $\Omega$-deformation,  turning on $N_i\neq 0$ would be introducing singularities in space-time which one would interpret in terms of surface operator insertions \cite{WittenGukov}.
In $\Omega$-background, one can turn on the vortex flux without inserting additional operators -- in fact, the only effect of the flux is to shift the effective values of the Coulomb moduli:\linebreak

$$
a_i \;\; \rightarrow \;\;  a_i-N_i \epsilon.
$$
In the $\Omega$-background, the 4d theory in the presence of  $N_i$ units of vortex flux  and with Coulomb branch scalar $a_i$ turned on is equivalent to the same theory without flux, but with $a_i$ shifted as above. In the $\Omega$-background, $a_i$ always appears in the combination \cite{NW}

$$
a_i+\epsilon wD_{i,w},
$$
where $D_w=\partial_w + A_{i,w}$ is the covariant derivative on the cigar $D$ with complex coordinate $w$. With $N_i$ units of flux on $D$, we have
$A_{i,w}= -N_i/w $, leading to the above result.\footnote{In \cite{NW} one proves that any flat gauge field on the punctured disk $D$, with origin $w=0$ deleted, preserves supersymmetry of the $\Omega$ background.}

For generic bare values of Coulomb branch moduli $a_i$,  and an arbitrary 4d ${\cal N}=2$ theory we have a single effective description of the 4d theory placed in 2d $\Omega$-background, with vortex flux turned -- this is the description we just gave. However, theories with baryonic Higgs branches have,
for a special value of $a_i$'s a {\it second} way to describe the same system.

We tune the bare values of Coulomb branch moduli $a_i$ so the 4d theory is at the root of the Higgs branch. There, the 4d theory has vortex solutions of charge $N_i$ -- this is the case upon varying FI D-term. The vortices are the non-abelian Nielsen-Olson vortices of \cite{HananyTong, HananyTong2}. We get a second $2d$ theory with ${\cal N}=(2,2)$ supersymmetry -- this is the theory on vortices themselves. In the theory on the vortex, the only effect of the $\Omega$-deformation is to give the scalar, parameterizing the position of the vortex in the $w$-plane, a twisted mass $\epsilon$.
The FI D-term that we needed to turn on to obtain the vortex solutions does not affect the F-terms of the 2d theory - it only changes the bare value of the gauge coupling, which is a D-term. The two descriptions must be the same in the IR, below the scale set by the 3d gauge coupling and $\epsilon$.

Having two descriptions of the same physics leads to a duality. What we cannot completely control in this analysis are the D-terms: the theory on the vortices may agree with the standard 2d gauge theory only up to D-term variation. So, we expect equivalence of the two theories up to D-term variation.

Example of this phenomenon is provided by ${\cal T}_{5d}$ and ${\cal V}_{3d}$, compactified on a circle. The gauge/vortex duality implies further that if we subject the theories to a full $\Omega$-background, their partition functions ${\cal Z}_{{\cal T}_{5d}}$ and ${\cal Z}_{{\cal V}_{3d}}$ will agree. Here, the parameter that we called $t$ in sections 2 and 3, with $t=e^{-R \epsilon}$, plays the crucial role in the duality. The second parameter, $q$, simply goes along for the ride -- it is necessary only when we compute partition functions.

\subsection{Proof of $A_n$-Triality, Part $ii.$}

We will now show that partition functions of 5d gauge theory ${\cal T}_{5d}$ on $M_{q,t}$ and its vortex 3d gauge theory ${\cal V}_{3d}$  on $M_q$ agree -- provided we tune the 5d Coulomb branch moduli to the root of the Higgs branch, and turn on flux. The fluxes shift the Coulomb branch moduli by the amount proportional to the vortex flux -- or the ranks of the 3d gauge group in the corresponding vacuum.

\begin{figure}[h!]
\begin{center}
\includegraphics[width=0.48\textwidth]{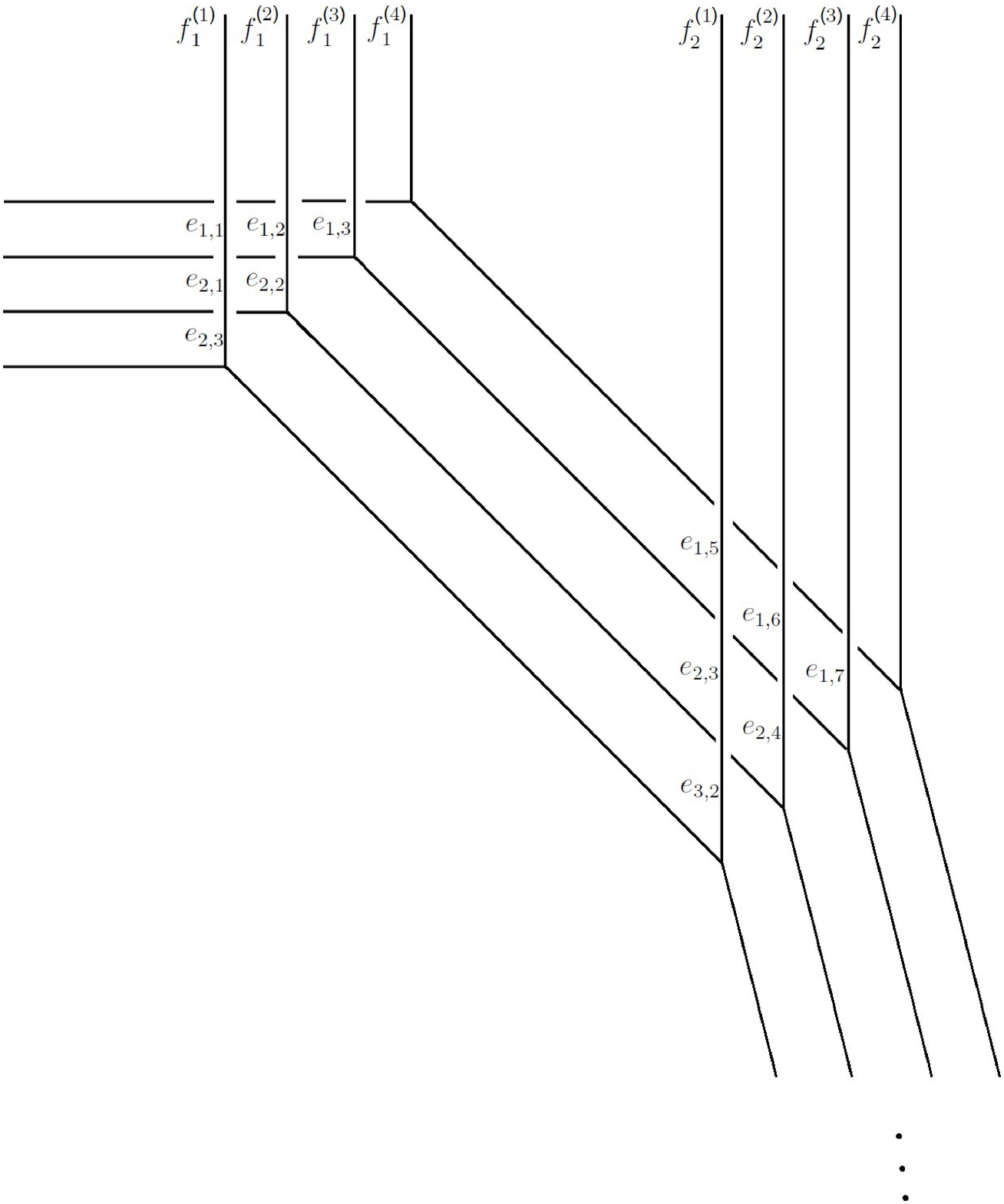}
\end{center}
\caption{A web diagram of the 5d gauge theory that corresponds to the spherical comb conformal block of $A_n$ (here $n = 3$) Toda theory with full punctures.}
\label{CFTversusQFTa}
\end{figure}

We will first start with the case when all punctures are full, with $\ell+2$ punctures in all. Then,  ${\cal T}_{5d}$ can be described by an $A_n$ quiver theory from section 2. The gauge group is a product of factors, $\prod_{a = 1}^{n} U\big(d_a\big)$, with ranks $d_a = (n + 1 - a)\ell$, and with $m_1 = (n+1){\ell}$ matter hypermultiplets in the fundamental representation of the leftmost factor $U\big(d_1\big) = U\big(n \ell\big)$. The theory can be conveniently represented by the web diagram on Fig. \ref{CFTversusQFTa}. The internal vertical lines are labeled by the (exponentials of the) Coulomb parameters, $e_{a,i}$, where $a = 1, \ldots, n$ and $i = 1, \ldots, d_a$. The external vertical lines are labeled by the   (exponentials of the) masses of the fundamentals \footnote{Note that the masses of the fundamentals are divided into $\ell$ groups -- each of these groups in the end will correspond to a full puncture on $C$. To better emphasize this fact, we borrow the notation $f^{(a)}_{i}$ for the masses from the 3d section: the index $i$ labels the groups, while the index $a$ distinguishes elements in a group.} $f^{(a)}_i$, where $a = 1, \ldots, n+1$ and $i = 1, \ldots, \ell$.

Let the Coulomb parameters take values
\begin{align}
e_{a,i} = v^{1-2a} \ t^{N_{a,i}} \ f^{(i \ {\rm mod} \ d_a)}_{i \div d_a}
\label{Truncation}
\end{align}
where $i \div d_a$ and $i \ {\rm mod} \ d_a$ are the quotient (rounded up) and remainder,\footnote{ This somewhat odd-looking rule is just a way to express, in a formula, the association of Coloumb parameters to the fundamentals shown geometrically on Fig. \ref{CFTversusQFTa}. } respectively, where $N_{a,i}$ are non-negative integers. Setting $N_{a,i}$ to zero corresponds to the root of the Higgs-branch (up to permutations).
The parameters $N_{a,i}$ determine the amount of shift of Coulomb branch moduli from the root of the Higgs branch: from $e_{a,i}=\exp(a_{a,i})$ to $e_{a,i}t^{N_{a,i}}=\exp(a_{a,i}-\epsilon N_{a,i})$. This is an effective description corresponding to staying at the point in the moduli space where the Coulomb and Higgs branches meet, and turning on vortex fluxes $N_{a,i}$ along the complex plane rotated by $t=\exp(-\epsilon)$.

Then, gauge/vortex duality gives ${\cal T}_{5d}$ has a dual description in terms of ${\cal V}_{3d}$, supported on the vortices wrapping $M_q$, in the vacuum where the 3d gauge group is broken as in \eqref{break}, namely from $G_{3d}$ to

\beq\label{breakagain}
 G_{3d}\quad \longrightarrow \quad \prod_{a=1}^n \prod_{i=1}^{d_{a}} U(N_{a,i}).
\eeq
We will now show that ${\cal Z}_{{\cal T}_{5d}}$, specialized to \eqref{Truncation} equals ${\cal Z}_{{\cal V}_{3d}}$, in the vacuum corresponding to \eqref{breakagain}.\footnote{The association of the ranks of the 3d gauge groups to vortex fluxes in 5d we give here is consistent with \eqref{Tong:2013iqa}.}

The partition function of ${\cal T}_{5d}$ is a sum
$$
Z_{{\cal T}_{5d}} = r_{5d} \sum\limits_{\{R\}} \ e^{\zeta \cdot R} \ I_{5d; \{ R \} }
$$
over $n(n+1) \ell/2$ partitions $R_{a,i}$, where $i = 1, \ldots, d_a$ and $a = 1, \ldots, n$. \linebreak
The summand is a product of Nekrasov factors,

\begin{align*}
I_{5d; \{ R \} } \ = \ &
\prod\limits_{a = 1}^{n+1} \prod\limits_{i = 1}^{l} \prod\limits_{j = 1}^{n l} \ {\cal N}_{\varnothing, R_{1,j}}\left( \frac{ v^{a-1} f^{(a)}_i }{ e_{1,j} } \right) \\
& \prod\limits_{a = 1}^{n} \prod\limits_{i = 1}^{d_a} \prod\limits_{j = 1}^{d_{a+1}} {\cal N}_{R_{a,i}, R_{a+1,j}}\left( \frac{ e_{a,i} }{ e_{a+1,j} } \right) \\
& \prod\limits_{a = 1}^{n} \prod\limits_{i,j = 1}^{d_a} {\cal N}_{R_{a,i}, R_{a,j}}\left( \frac{ e_{a,i} }{ e_{a,j} } \right)^{-1} \\
& \prod\limits_{a = 1}^{n} \prod\limits_{i = 1}^{d_a} \big(T_{R_{a,i}}\big)^{\ell}
\end{align*}
where the first, second, third lines represent, respectively, are the contributions of the fundamental, bifundamental, and gauge vector multiplets, and the last line is the contribution of the 5d Chern-Simons terms. The building blocks ${\cal N}_{RP}$ and $T_R$ were defined previously in s.2.4.

The starting step of the proof is the observation, that Nekrasov factor ${\cal N}_{RP}\big( v^2 t^{-N} \big)$ with two indices $R,P$ and a non-negative integer $N$ vanishes, unless $l(P) \leq l(R) + N$. This observation is a simple algebraic corollary of the explicit formulas for Nekrasov factors, and we take it for granted. There are enough Nekrasov factors in the numerator of $I_{5d}$ to imply that, if the Coloumb parameters $e_{a,i}$ are chosen according to the \eqref{Truncation} rule, then each partition $R_{a,i}$ has no more than $N_{a,i}$ rows, i.e., $l(R_{a,i}) \leq N_{a,i}$.

The next step of the proof is to rewrite the Nekrasov functions, which are \emph{a priori} defined as infinite double products

\begin{align*}
{\cal N}_{R,P}(Q) = \prod\limits_{i = 1}^{\infty} \prod\limits_{j = 1}^{\infty}
\dfrac{\varphi\big( Q q^{R_i-P_j} t^{\rho_i-\rho_j + 1} \big)}{\varphi\big( Q q^{R_i-P_j} t^{\rho_i-\rho_j} \big)} \
\dfrac{\varphi\big( Q t^{\rho_i-\rho_j} \big)}{\varphi\big( Q t^{\rho_i-\rho_j + 1} \big)}
\end{align*}
in terms of finite products, bounded by $l(R)$ and $l(P)$. Since all the partitions now have finite length, this is possible to do: one just needs to break down the above infinite product over $(i,j)$ into three parts: the product over $0 \leq i \leq l(R), 0 \leq j \leq l(P)$; the product over $0 \leq i \leq l(R), j \geq l(P)$; the product over $i \geq l(R), 0 \leq j \leq l(P)$. The latter two products are formally infinite, but enjoy telescoping and hence are, in fact, finite products. Applying this for the Nekrasov factor ${\cal N}_{R,P}(Q)$ where partitions $R,P$ have corresponding Coloumb parameters $e_1,e_2$ and lengths $N_1,N_2$, we obtain

\begin{align*}
{\cal N}_{RP}\Big( \frac{e_1}{e_2} \Big) = \ &
\prod\limits_{i = 1}^{N_1} \prod\limits_{j = 1}^{N_2} \dfrac{\varphi\big( \frac{e_1}{e_2} q^{R_i-P_j} t^{\rho_i-\rho_j + 1} \big)}{\varphi\big( \frac{e_1}{e_2} q^{R_i-P_j} t^{\rho_i-\rho_j} \big)} \ \dfrac{\varphi\big( \frac{e_1}{e_2} t^{\rho_i-\rho_j} \big)}{\varphi\big( \frac{e_1}{e_2} t^{\rho_i-\rho_j + 1} \big)} \\
& N_{R, \varnothing}\Big( t^{N_2} \frac{e_1}{e_2} \Big) N_{\varnothing, P}\Big( t^{- N_1} \frac{e_1}{e_2} \Big)
\end{align*}
The right hand side, being a ratio of products of quantum dilogarithms taken at values shifted by $t$, is very much reminiscent of $\Phi_V$, the vector multiplet contribution to the index of 3d theory that we encountered in section 3. To make this observation precise, let us apply this formula to rewrite the contribution of the 5d vector multiplets as

$$
\prod\limits_{a = 1}^{n} \prod\limits_{i,j = 1}^{d_a} {\cal N}_{R_{a,i}, R_{a,j}}\left( \frac{ e_{a,i} }{ e_{a,j} } \right)^{-1} = \prod\limits_{a = 1}^{n} \dfrac{\Phi_{V_a}\big(x^{(a)}\big)}{\Phi_{V_a}\big(x_{\varnothing}^{(a)}\big)} \cdot V_{\rm vect}
$$
where $\Phi_{V_a}$ is the contribution of the 3d vector multiplet corresponding to a gauge group of rank $N^{(a)}$ to the index of the 3d theory, evaluated at positions 

\begin{align}
e^{-x^{(a)}} = \{ v^{2a} e_{a,i} t^{\rho} q^{R_{a,i}} \}_{i = 1, \ldots, d_a}
\label{xpos}
\end{align} 
and $x^{(a)}_{\varnothing}$ is the specialization of $x^{(a)}$ to empty partition $R_{a,i} = \varnothing$, while $V_{\rm vect}$ stands for all the remaining factors, which we leave untouched for now:

$$
V_{\rm vect} = \prod\limits_{a = 1}^{n} \prod\limits_{i,j = 1}^{d_a}{\cal  N}_{R_{a,i}, \varnothing}\Big( t^{N_{a,j}} \frac{e_{a,i}}{e_{a,j}} \Big)^{-1}
 {\cal N}_{\varnothing, R_{a,j}}\Big( t^{- N_{a,i}} \frac{e_{a,i}}{e_{a,j}} \Big)^{-1}
$$
In complete analogy, the contribution of 5d bifundamentals takes form

$$
\prod\limits_{a = 1}^{n} \prod\limits_{i = 1}^{d_a} \prod\limits_{j = 1}^{d_{a+1}} {\cal N}_{R_{a,i}, R_{a+1,j}}\left( \frac{ e_{a,i} }{ e_{a+1,j} } \right) = \prod\limits_{a = 1}^{n} \dfrac{\Phi_{H^{a,a+1}}\big(x^{(a)} ,x^{(a+1)} \big)}{\Phi_{H^{a,a+1}}\big(x^{(a)}_{\varnothing},x^{(a+1)}_{\varnothing}\big)} \cdot V_{\rm bifund}
$$
where $\Phi_{H^{a,a+1}}$ is the contribution of the 3d bifundamental multiplet corresponding to a gauge group of rank $N_{a}$ to the index of the 3d theory, and $V_{\rm bifund}$ stands for all the remaining factors, which we similarly leave for now:

$$
V_{\rm bifund} = \prod\limits_{a = 1}^{n} \prod\limits_{i = 1}^{d_a} \prod\limits_{j = 1}^{d_{a+1}} N_{R_{a,i},\varnothing}\Big( t^{N_{a+1,j}} \frac{e_{a,i}}{e_{a+1,j}} \Big) N_{\varnothing, R_{a+1,j}}\Big( t^{- N_{a,i}} \frac{e_{a,i}}{e_{a+1,j}} \Big)
$$
At this point, putting all expressions together, we uncover

$$
Z_{{\cal T}_{5d}}= r_{5d} \sum\limits_{\{R\}} \ e^{\zeta \cdot R} \ \left( \prod\limits_{a = 1}^{n} \dfrac{\Phi_{V_a}\big(x^{(a)}\big)}{\Phi_{V_a}\big(x_{\varnothing}^{(a)}\big)} \dfrac{\Phi_{H^{a,a+1}}\big(x^{(a)} ,x^{(a+1)} \big)}{\Phi_{H^{a,a+1}}\big(x^{(a)}_{\varnothing},x^{(a+1)}_{\varnothing}\big)} \right) \cdot V_{\rm vect} V_{\rm bifund} V_{\rm fund} V_{\rm CS} 
$$
where $V_{\rm fund}$ stands for the contribution of fundamentals

$$
V_{\rm fund} = \prod\limits_{a = 1}^{n+1} \prod\limits_{i = 1}^{l} \prod\limits_{j = 1}^{n l} \ {\cal N}_{\varnothing, R_{1,j}}\left( \frac{ v^{a-1} f^{(a)}_i }{ e_{1,j} } \right)
$$
and $V_{\rm CS} = \prod_{a,i} T_{R_{a,i}}^\ell$ is the 5d Chern-Simons contribution.

The product $V_{\rm vect} V_{\rm bifund} V_{\rm fund} V_{\rm CS}$ may appear to have a lot of factors, however, there are many cancellations, implied by the identifications (\ref{Truncation}). After the cancellations are fully accounted for, this product takes form\footnote{Note, that the 5d Chern-Simons terms are completely cancelled out in the process.}

$$
V_{\rm vect} V_{\rm bifund} V_{\rm fund} V_{\rm CS} = \prod\limits_{a = 1}^{n} \dfrac{\Phi_{H_a}\left(x^{(a)}\right)}{\Phi_{H_a}\left(x^{(a)}_{\varnothing}\right)}
$$
where $\Phi_{H_a}$ is the contribution of 3d fundamentals to the index of the 3d theory, with $f^{(a)}_i$ we introduced earlier being the (inverses of) masses of 3d fundamentals. Taking this into account, we arrive at the following expression for the instanton partition function, which is highly reminiscent of the 3d index that we described in section 2:

$$
Z_{{\cal T}_{5d}} = r_{5d}\sum\limits_{\{R\}} \ e^{\zeta \cdot R} \ I_{5d}, \ \ \ I_{5d} = \prod\limits_{a = 1}^{n}
\dfrac{\Phi_{V_a}\left(x^{(a)}\right)}{\Phi_{V_a}\left(x_{\varnothing}^{(a)}\right)} \dfrac{\Phi_{H^{a,a+1}}\big(x^{(a)} ,x^{(a+1)} \big)}{\Phi_{H^{a,a+1}}\left(x^{(a)}_{\varnothing},x^{(a+1)}_{\varnothing}\right)} \dfrac{\Phi_{H_a}\big(x^{(a)}\big)}{\Phi_{H_a}\big(x^{(a)}_{\varnothing}\big)}
$$
So far we have been pedantically keeping track of the normalization factors at $x = x_{\varnothing}$. This is done for a reason: both the numerators and the denominators do not make sense on their own, because they have poles precisely at positions (\ref{xpos}) for non-negative integer values of $R_{a,i}$. The ratios, however, are finite and coincide with the ratios of residues:

$$
I_{5d}
= {\rm res}_{\varnothing}^{-1} \cdot {\rm res}_{R} \left( \prod\limits_{a = 1}^{n} e^{ \zeta_a \,{\rm Tr}\, x^{(a)}/\hbar} \Phi_{V_a}\big(x^{(a)}\big) \Phi_{H^{a,a+1}}\big(x^{(a)} ,x^{(a+1)} \big) \Phi_{H_a}\big(x^{(a)}\big) \right)
$$
Here ${\rm res}_{R}$ stands for the residue of the bracketed expression at the point (\ref{xpos}) and ${\rm res}_{\varnothing}$ is the residue of the same expression at the point $x^{(a)} = x^{(a)}_{\varnothing}$. Hence $Z_{5d}$ is equal, up to the overall proportionality factor, to the sum of residues of the 3d index, i.e. to that index itself:

$$
Z_{{\cal T}_{5d}} = \dfrac{r_{5d}}{{\rm res}_{\varnothing}} \sum\limits_{\{R\}} \ {\rm res}_{R} \left( \prod\limits_{a = 1}^{n} \Phi_{V_a}\big(x^{(a)}\big) \Phi_{H^{a,a+1}}\big(x^{(a)} ,x^{(a+1)} \big) \Phi_{H_a}\big(x^{(a)}\big) \right) =
$$
\begin{align}
= \dfrac{r_{5d}}{{\rm res}_{\varnothing}} \oint dX_1 \ldots dX_n \ \prod\limits_{a = 1}^{n} \Phi_{V_a}\big(x^{(a)}\big) \Phi_{H^{a,a+1}}\big(x^{(a)} ,x^{(a+1)} \big) \Phi_{H_a}\big(x^{(a)}\big)
\label{DFfinal}
\end{align}
This completes the proof. \smallskip\\

Note that the constant of proportionality $r_{5d}/{\rm res}_{\varnothing}$ is not one only because, in this derivation, we have chosen to consider only the purely 3d gauge theory contributions to ${\cal Z}_{{\cal V}_{3d}}$.  If we included into consideration of section 3 also the contribution of the bulk (corresponding to the 5d partition function on the Higgs branch, where the vortices and ${\cal V}_{3d}$ live), then the equality between the 5d gauge theory partition function ${\cal Z}_{{\cal T}_{5d}}$ and the ${\cal Z}_{{\cal V}_{3d}}$ would be exact, without any extra proportionality constant. We have checked that this is indeed the case.

\paragraph{A digression on residues.} We have chosen to present this proof by proceeding from the 5d Nekrasov partition function ${\cal Z}_{{\cal T}_{5d}}$ to the 3d gauge partition function ${\cal Z}_{{\cal V}_{3d}}$. However, it is of course possible to go the other way,
starting with ${\cal Z}_{{\cal V}_{3d}}$ and showing how do the Nekrasov factors appear as its residues. We will present this argument here but, for the sake of transparency, restrict ourselves to the case $n = 2, \ell = 1$, which corresponds to the $A_2$-Toda theory on a sphere with three punctures. In that case, there are two groups of variables $x^{(1)}, x^{(2)}$ (with $N_1, N_2$ elements each), that we will call $X,Y$ in what follows, and

$$
Z_{{\cal V}_{3d}} = \oint dX \oint dY \ \Phi_{V_1}(X) \Phi_{V_2}(Y) \ \Phi_{H_1}(X) \Phi_{H_2}(Y) \ \Phi_{H^{12}}(X,Y)
$$
is the 3d index, with potentials
$$
\Phi_{H_1}(X) = \prod\limits_{x \in X} \dfrac{\varphi\left( q^{-\alpha_{1}} w / x \right)}{\varphi\left( v^{-1} q^{-\alpha_{2}} w / x \right)} , \ \ \ \Phi_{H_2}(Y) = \prod\limits_{y \in Y} \dfrac{\varphi\left( q^{-\alpha_{2}} w / y \right)}{\varphi\left( v^{-1} q^{-\alpha_{3}} w / y \right)}
$$
To fully specify the integral, one should fix the contours of integration in it. As explained in section 3, there are two contours to choose from. The most obvious choice is to encircle points 0 and $v^{-1} q^{-\alpha_{2}} w$ for $X$ (resp. points 0 and $v^{-1} q^{-\alpha_{3}} w$ for $Y$). Another, less obvious, is to choose a contour for the variable $X_i$ that encircles points 0 and $v Y_i$. This possibility is essentially new compared to the $A_1$ case that we previously considered in \cite{AHKS}, since it is dictated by the form of the interaction $\Phi_{H^{12}}$ which in the $A_1$ case never appears.

Because two options for the choice of contours exist, each group of variables splits into two subsets -- this splitting corresponds to breaking of the $G_{3d}$ gauge symmetry. Let us start by first integrating out $X$, and then $Y$. For $X$, there are two non-equivalent contours available. Let $X_1 \subset X$ be those $X$-variables, for which the contour encircles the points 0 and $X = v^{-1} Y$. Note that this requires a choice of (arbitrary, thanks to permutation symmetry) subset $Y_1 \subset Y$. The number of elements in $X_1$ and in $Y_1$ should be one and the same, let us denote it $M_1$. The remaining variables in $X_2$ get integrated over a contour that encircles 0 and $v^{-1} q^{-\alpha_{2}} w$, the number of such is $M_2$, with $N_1=M_1+M_2$. After we integrated over the $X$-variables and proceed to integrating over the $Y$-variables, there is only a single contour available, namely, that encircles 0 and $v^{-1} q^{-\alpha_{3}} w$, the number of such is $N_2 =M_1+M_3$.

\begin{figure}[h!]
\begin{center}
\includegraphics[width=0.4737\textwidth]{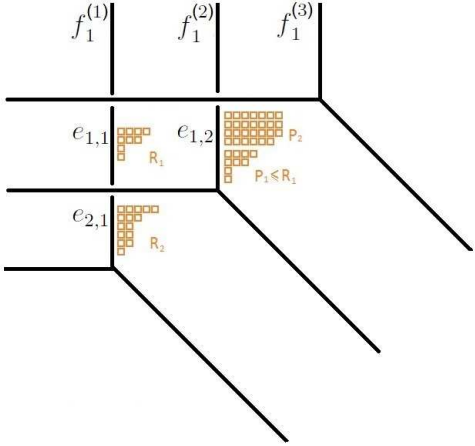}
\end{center}
\caption{The finer structure of poles. }
\end{figure}

Having specified the contours of integration, we can evaluate the integral as a residue sum over positions of poles. The poles that occur for $X_2$ inside of the contour, encircling the points $0$ and $v^{-1} q^{-\alpha_{2}} w$, come from two distinct sources: the poles of the vector contribution $\Phi_{V_1}(X)$, and the poles of the flavor contribution $\Phi_{H_1}(X)$. As explained in more detail in \cite{AHKS}, the former has poles proportional to a vector $t^{M_2 + \rho} q^{R_2}$, the latter fixes the proportionality factor to be at the pole of the potential, i.e. at $v^{-1} q^{-\alpha_{2}} w$. We thus get poles at positions
$$
v^{-1} q^{-\alpha_{2}} w \ \big( t^{M_2 + \rho} q^{R_2} \big)_i = v e_{1,2} \big( t^{\rho} q^{R_2} \big)_i
$$
where $R_2$ is a partition of length $N_2$. Similarly, for $Y$ we get poles at positions
$$
v^{-1} q^{-\alpha_{3}} w \ \big( t^{M_1 + M_3 + \rho} q^{R_2} \big)_i = v^2 e_{2,1} \big( t^{\rho} q^{P} \big)_i
$$
where $P$ is some partition of length $M_1+M_3$ that consists two parts, $P = (P_2, P_1)$, corresponding to $Y_2$ and $Y_1$, of lengths $M_3$ and $M_1$, respectively.

Finally, for $X_1$ the vector multiplet contribution does not affect poles at all, because of a cancellation between $\Phi_{V_1}$ and $\Phi_{H^{12}}$ from the denominator. All the poles are determined now by $\Phi_{H^{12}}(X_1,X_2)$ and situated at
$$
v^{-1} (Y_1)_i \ \big( q^{H} \big)_i = v e_{1,1} \big( t^{\rho} q^{R_1} \big)_i
$$
where $H$ and $R_1 = H + P_1$ are partitions of length $M_1$. One can see that the poles for the $a$-th group of variables are at positions $\{ v^a e_{a,i} t^{\rho} q^{R_{a,i}} \}_{i = 1, \ldots, d_a}$, in accordance with the first part of this section. Computation of the residues at these poles and matching them with Nekrasov factors would replicate the first part of this section. This completes the proof.

Interestingly, this approach -- starting from the 3d gauge theory side -- reveals a finer structure in the Nekrasov series expansion of the instanton partition function. Namely, as one can see from the above formulas, the partitions $R_1$ and $P_1$ are not entirely independent: they satisfy a relation $R_1 = H + P_1$ and hence an inequality $R_1 \geq P_1$. Graphically, this means that the Young diagram associated to the Coloumb parameter $e_{1,1}$ is bounded by the (second part of) the Young diagram associated to the Coloumb parameter $e_{2,1}$. One can check that Nekrasov factors indeed have this property (vanish, unless $R_1 \geq P_1$), as one would expect from the residues of the 3d partition function, in the  $A_2$ case.

\subsection{Generalization to other ${\cal T}_{5d}$ gauge theories}

We just demonstrated explicitly that ${\cal Z}_{{\cal V}_{3d}}$ and ${\cal Z}_{{\cal T}_{5d}}$ agree in the case when all punctures are full.  Here we explain how this extends to other cases, corresponding to two full punctures, at $z=0, \infty$ and arbitrary punctures elsewhere. This corresponds to ${\cal T}_{5d}$ which is a general quiver gauge theory, with a gauge group $\prod_{a = 1}^{n} U\big(d_a\big)$ and arbitrary number $m_a$ of matter hypermultiplets in the fundamental representation of the $a$-th factor, subject to constraints \eqref{conf}.

\begin{figure}[h!]
\begin{center}
\includegraphics[width=0.2\textwidth]{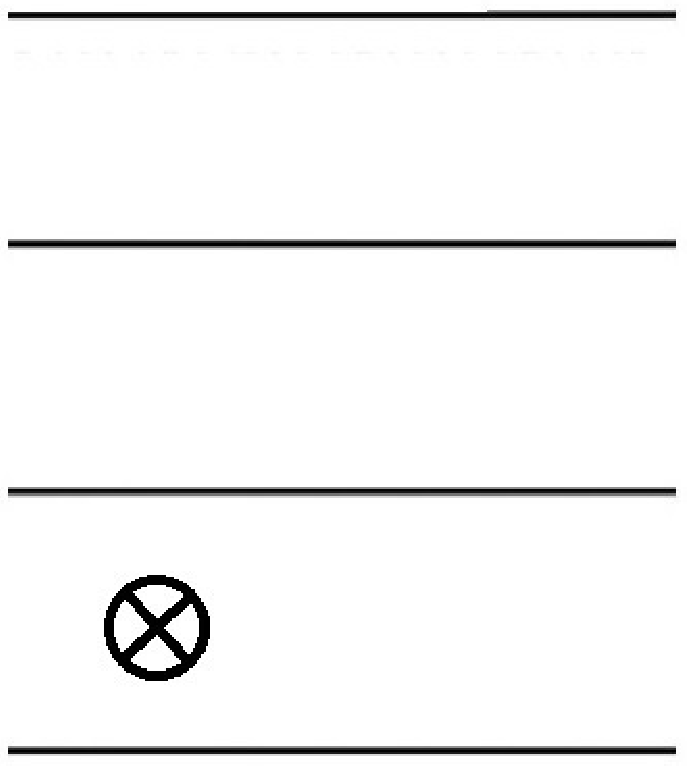}
\ \ \ \ \ \ \ \ \ \
\includegraphics[width=0.2\textwidth]{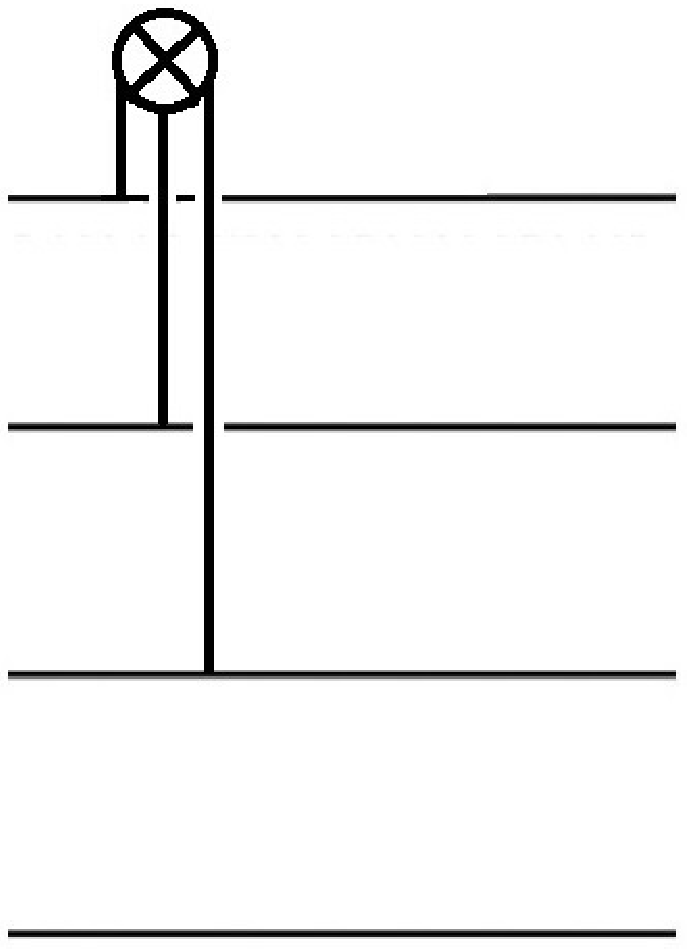}
\end{center}
\caption{A Hanany-Witten move.}
\label{HananyWitten}
\end{figure}

The instanton partition function of this more general theory is a sum
$$
Z_{{\cal T}_{5d}} = \sum\limits_{\{R\}} I_{5d}
$$
over $d_1 + \ldots + d_n$ partitions $R_{a,i}$, where $i = 1, \ldots, d_a$ and $a = 1, \ldots, n$. \linebreak
The summand is a product of Nekrasov factors,
\begin{align*}
I_{5d} \ = \ &
\prod\limits_{a = 1}^{n} \prod\limits_{i = 1}^{m_a} \prod\limits_{j = 1}^{d_a} \ {\cal N}_{\varnothing, R_{a,j}}\left( \frac{ v f_{a,i} }{ e_{a,j} } \right) \\
& \prod\limits_{a = 1}^{n} \prod\limits_{i = 1}^{d_a} \prod\limits_{j = 1}^{d_{a+1}} {\cal N}_{R_{a,i}, R_{a+1,j}}\left( \frac{ e_{a,i} }{ e_{a+1,j} } \right) \\
& \prod\limits_{a = 1}^{n} \prod\limits_{i,j = 1}^{d_a} {\cal N}_{R_{a,i}, R_{a,j}}\left( \frac{ e_{a,i} }{ e_{a,j} } \right)^{-1} \\
& \prod\limits_{a = 1}^{n} \prod\limits_{i = 1}^{d_a} \big(T_{R_{a,i}}\big)^{d_a - d_{a+1}}
\end{align*}
where the first, second, third lines represent, respectively, the contributions of the fundamental, bifundamental, and gauge vector multiplets, and the last line is the contribution of the 5d Chern-Simons terms. The previous section corresponds to a particular case of $\big(m_1, \ldots, m_n\big) = \big( (n+1)\ell, 0, \ldots, 0 \big)$.

To find the corresponding ${\cal Z}_{{\cal V}_{3d}}$ one may, just as before, rewrite $I_{5d}$ as a residue of a certain product of quantum dilogarithms -- the 3d integrand $I_{3d}$. Doing so directly, however, is not the most efficient way to proceed. It is more convenient to represent ${\cal Z}_{{\cal T}_{5d}}$ as a specialization of the theory we considered in the previous section. By doing so, we reduce the problem to the previously solved case.

We will prove the following statement:

\paragraph{Proposition.} The partition function of ${\cal T}_{5d}$ can be obtained from the partition function of another theory ${\cal T}^{{\rm full}}_{5d}$, with {\fontsize{9pt}{0pt}{$(m^{{\rm full}}_1, \ldots, m^{{\rm full}}_n) = \big((n+1)\ell, 0, \ldots, 0 \big)$}}. Upon specialization of the moduli of ${\cal T}^{{\rm full}}_{5d}$, the partition functions equal,
$${\cal Z}_{{\cal T}_{5d}} = {\cal Z}_{{\cal T}_{5d}^{{\rm full}}}.$$
\paragraph{Proof.} We will prove this recursively in fundamental hypermultiplets.

The theories ${\cal T}_{5d}$ and ${\cal T}_{5d}^{{\rm full}}$, where we specialize the moduli of the latter, are related by a sequence of Hanany-Witten \cite{Hanany:1996ie} moves, see Fig. \ref{HananyWitten}. The elementary step of recursion is to relate the partition function of ${{\cal T}_{5d}}$ to that of another theory ${{\cal T}_{5d}}'$, with special moduli. We show that, upon specialization, ${\cal Z} = {\cal Z}'$. To obtain ${\cal T}'$ from ${\cal T}$ we choose a fundamental for $k$-th gauge group of ${\cal T}$, remove it and add $k$ more fundamentals for the $1$-st gauge group. This means that in ${\cal T}'$ we have $m^{'}_1 = m_1 + k$, $m^{\prime}_k = m_k - 1$ and $m^{\prime}_a = m_a$ otherwise. To satisfy \eqref{conf}, we must change the ranks of the gauge groups. For $a<k$, we increase the rank of the $a$-th gauge group in the quiver by $k - a$: i.e. the rank $d^{\prime}_a = d_a + k - a$ for $a < k$ and $d^{\prime}_a = d_a$ otherwise. The new Coulomb parameters, say, $e^{\prime}_{a,i}$ with $i = 1, \ldots, k - a$ and $a = 1, \ldots, k-1$, are related to the mass of the removed fundamental via $e^{\prime}_{a,i} = v^{1-2a-2i} f$, and the masses of the new fundamentals, say, $f^{\prime}_{1,1}, \ldots, f^{\prime}_{1,k}$ are related to the mass of the removed one via $f^{\prime}_{1,i} = v^{-2i} f$.

What we need to prove is ${\cal Z} = {\cal Z}^{\prime}$. This is a direct calculation via
$$
{\cal Z} = \sum\limits_{\{R\}} I_{\{R\}}, \ \ \ \ \ {\cal Z}^{\prime} = \sum\limits_{\{R\}} I^{\prime}_{\{R\}}
$$
Note that, at a first thought, $Z^{\prime}$ is a sum over a bigger number of partitions than $Z$. Indeed, the partitions are in one-to-one correspondence with the Coloumb parameters, and we added some. However, neither of these extra partitions actually contributes to the partition function, for the following reason. Recall that, in the previous section, we noted that Nekrasov factor ${\cal N}_{PR}(v^2)$ vanishes unless the diagram $R$ is contained in the diagram $P$. All the added parameters (both the new Coulomb parameters and the masses of new fundamentals) are nearly equal to each other and to $f$, as shown on Fig. \ref{HananyWitten}. The only difference is in the $v$-shifts, and these $v$-shifts are precisely such that the arguments of Nekrasov factors for any pair of partitions associated to the newly added lines is $v^2$. This implies a chain of inclusions which forces each of these newly added partitions to be contained in an empty partition $\varnothing$, which is the one associated to the external lines. Since the only partition contained in $\varnothing$ is $\varnothing$ itself, we see that all partitions associated to the newly added lines are empty.

Hence, $Z$ and $Z^{\prime}$ are actually sums over the same set of partitions, and it remains just to prove that $I = I^{\prime}$. This is easy to do: indeed,

\begin{align*}
\dfrac{ I^{\prime}_{\{R\}} }{ I_{\{R\}} } & = \prod\limits_{a = 1}^{k-1} \prod\limits_{i = 1}^{d_a} \ T_{R_{a,i}} \ \prod\limits_{a = 1}^{n} \prod\limits_{i = 1}^{d_a} \dfrac{\prod\limits_{j = 1}^{k-a+1} {\cal N}_{\varnothing, R_{a,i}}\left(\dfrac{e^{\prime}_{(a-1),j}}{e_{a,i}}\right) \prod\limits_{j = 1}^{k-a-1} {\cal N}_{R_{a,i},\varnothing}\left(\dfrac{e_{a,i}}{e^{\prime}_{a+1,j}}\right)}{\prod\limits_{j = 1}^{k-a}{\cal N}_{\varnothing, R_{a,i}}\left(\dfrac{e^{\prime}_{a,j}}{e_{a,i}}\right) \prod\limits_{j = 1}^{k-a} {\cal N}_{R_{a,i},\varnothing}\left(\dfrac{e_{a,i}}{e^{\prime}_{a,j}}\right)} = 
\end{align*}
\begin{align*}
\prod\limits_{a = 1}^{n} \prod\limits_{i = 1}^{d_a} \dfrac{\prod\limits_{j = 1}^{k-a+1} {\cal N}_{\varnothing, R_{a,i}}\left(\dfrac{e^{\prime}_{(a-1),j}}{e_{a,i}}\right) \prod\limits_{j = 1}^{k-a-1} {\cal N}_{\varnothing, R_{a,i}}\left(v^2 \dfrac{e^{\prime}_{a+1,j}}{e_{a,i}}\right)}{ \prod\limits_{j = 1}^{k-a} {\cal N}_{\varnothing, R_{a,i}}\left(\dfrac{e^{\prime}_{a,j}}{e_{a,i}}\right) \prod\limits_{j = 1}^{k-a} {\cal N}_{\varnothing,R_{a,i}}\left(v^2 \dfrac{e^{\prime}_{a,j}}{e_{a,i}}\right)} = 
\end{align*}
\begin{align}
\hspace{-2ex} = \prod\limits_{a = 1}^{n} \prod\limits_{i = 1}^{d_a} \dfrac{ \prod\limits_{j = 1}^{k-a+1} {\cal N}_{\varnothing, R_{a,i}}\left(\dfrac{v^{-2j-2a+2}}{e_{a,i}}\right) \prod\limits_{j = 1}^{k-a-1} {\cal N}_{\varnothing, R_{a,i}}\left(\dfrac{v^{-2a-2j}}{e_{a,i}}\right)}{ \prod\limits_{j = 1}^{k-a} {\cal N}_{\varnothing, R_{a,i}}\left(\dfrac{v^{-2a-2j}}{e_{a,i}}\right) \prod\limits_{j = 1}^{k-a} {\cal N}_{\varnothing,R_{a,i}}\left(\dfrac{v^{-2j-2a+2}}{e_{a,i}}\right)} = 1
\end{align}
\smallskip\\
where to pass from the first line to the second we used the property of Nekrasov factors ${\cal N}_{\varnothing,R}( Q ) = T_R {\cal N}_{R,\varnothing}( v^2 Q^{-1} )$ where $T_R$ are the framing factors, and to pass from the second line to the third we used the specialized values of the new Coulomb parameters. Note, how the framing factors cancel out in the process. Note also, that in the above we employed $e^{\prime}_{(0),j} = v f^{\prime}_{(1),j}$, which shortens the computation allowing to treat the contributions of fundamentals as a particular case of the contribution of bifundamentals. Hence $I = I^{\prime}$, and $Z = Z^{\prime}$.

This completes the proof of the recursive step. To prove the proposition, simply apply the above to each fundamental multiplet in ${{\cal T}_{5d}}$ one step at a time. The ultimate result is a certain specialization of a theory with $(m^{{\rm full}}_1, \ldots, m^{{\rm full}}_n) = \big((n+1)\ell, 0, \ldots, 0 \big)$ fundamentals, whose partition function equals ${\cal Z}_{{\cal T}_{5d}}$. The web diagram of this theory is Fig.(\ref{GaugeWeb}). The proposition in fact proves that partition functions of theories related by Hanany-Witten moves are the same.

\section*{Acknowledgments}
 We are grateful to Hiraku Nakajima, Andrei Okounkov, Joerg Teschner, David Tong and Cumrun Vafa for valuable discussions. The research of M.A., N. H. and S. S is supported in part by the Berkeley Center for Theoretical Physics, by the National Science Foundation (award number 0855653), by the Institute for the Physics and Mathematics of the Universe, and by the US Department of Energy under Contract DE-AC02-05CH11231. The work of S. S is partly supported by grant RFBR 13-02-00478, grant for support of scientic schools Nsh-3349.2012.2 and government contract 8207.

\bibliographystyle{utcaps}
\bibliography{myrefs4}

\providecommand{\href}[2]{#2}\begingroup\raggedright\begin{thebibliography}{10}

\bibitem{AGT}
L.~F. Alday, D.~Gaiotto, and Y.~Tachikawa, ``{Liouville Correlation Functions
  from Four-dimensional Gauge Theories},''
  \href{http://dx.doi.org/10.1007/s11005-010-0369-5}{{\em Lett.Math.Phys.} {\bf
  91} (2010)  167--197},
\href{http://arxiv.org/abs/0906.3219}{{\tt arXiv:0906.3219 [hep-th]}}.

\bibitem{G2}
D.~Gaiotto, ``{N=2 dualities},''
  \href{http://dx.doi.org/10.1007/JHEP08(2012)034}{{\em JHEP} {\bf 1208} (2012)
   034},
\href{http://arxiv.org/abs/0904.2715}{{\tt arXiv:0904.2715 [hep-th]}}.

\bibitem{Gaiotto:2009hg}
D.~Gaiotto, G.~W. Moore, and A.~Neitzke, ``{Wall-crossing, Hitchin Systems, and
  the WKB Approximation},''
\href{http://arxiv.org/abs/0907.3987}{{\tt arXiv:0907.3987 [hep-th]}}.

\bibitem{DH1}
H.-Y. Chen, N.~Dorey, T.~J. Hollowood, and S.~Lee, ``{A New 2d/4d Duality via
  Integrability},'' \href{http://dx.doi.org/10.1007/JHEP09(2011)040}{{\em JHEP}
  {\bf 1109} (2011)  040},
\href{http://arxiv.org/abs/1104.3021}{{\tt arXiv:1104.3021 [hep-th]}}.

\bibitem{DH2}
N.~Dorey, S.~Lee, and T.~J. Hollowood, ``{Quantization of Integrable Systems
  and a 2d/4d Duality},'' \href{http://dx.doi.org/10.1007/JHEP10(2011)077}{{\em
  JHEP} {\bf 1110} (2011)  077},
\href{http://arxiv.org/abs/1103.5726}{{\tt arXiv:1103.5726 [hep-th]}}.

\bibitem{SimonsTalk}
M.~Aganagic, ``{M-theory, Large $N$ Duality and the Dynamics of Vortices},''
  {\em Talk at 11th Simons Summer Workshop at SCGP} (2013)  .

\bibitem{Dorey}
N.~Dorey, ``{The BPS spectra of two-dimensional supersymmetric gauge theories
  with twisted mass terms},'' {\em JHEP} {\bf 9811} (1998)  005,
\href{http://arxiv.org/abs/hep-th/9806056}{{\tt arXiv:hep-th/9806056
  [hep-th]}}.

\bibitem{DHT}
N.~Dorey, T.~J. Hollowood, and D.~Tong, ``{The BPS spectra of gauge theories in
  two-dimensions and four-dimensions},'' {\em JHEP} {\bf 9905} (1999)  006,
\href{http://arxiv.org/abs/hep-th/9902134}{{\tt arXiv:hep-th/9902134
  [hep-th]}}.

\bibitem{HananyTong}
A.~Hanany and D.~Tong, ``{Vortices, instantons and branes},'' {\em JHEP} {\bf
  0307} (2003)  037,
\href{http://arxiv.org/abs/hep-th/0306150}{{\tt arXiv:hep-th/0306150
  [hep-th]}}.

\bibitem{HananyTong2}
A.~Hanany and D.~Tong, ``{Vortex strings and four-dimensional gauge
  dynamics},'' \href{http://dx.doi.org/10.1088/1126-6708/2004/04/066}{{\em
  JHEP} {\bf 0404} (2004)  066},
\href{http://arxiv.org/abs/hep-th/0403158}{{\tt arXiv:hep-th/0403158
  [hep-th]}}.

\bibitem{Shifman:2004dr}
M.~Shifman and A.~Yung, ``{NonAbelian string junctions as confined
  monopoles},'' \href{http://dx.doi.org/10.1103/PhysRevD.70.045004}{{\em
  Phys.Rev.} {\bf D70} (2004)  045004},
\href{http://arxiv.org/abs/hep-th/0403149}{{\tt arXiv:hep-th/0403149
  [hep-th]}}.

\bibitem{Dotsenko:1984nm}
V.~Dotsenko and V.~Fateev, ``{Conformal Algebra and Multipoint Correlation
  Functions in Two-Dimensional Statistical Models},''
\href{http://dx.doi.org/10.1016/0550-3213(84)90269-4}{{\em Nucl.Phys.} {\bf
  B240} (1984)  312}.

\bibitem{AHKS}
M.~Aganagic, N.~Haouzi, C.~Kozcaz, and S.~Shakirov, ``{Gauge/Liouville
  Triality},''
\href{http://arxiv.org/abs/1309.1687}{{\tt arXiv:1309.1687 [hep-th]}}.

\bibitem{GV}
R.~Gopakumar and C.~Vafa, ``{On the gauge theory/geometry correspondence},''
  {\em Adv. Theor. Math. Phys.} {\bf 3} (1999)  1415--1443,
\href{http://arxiv.org/abs/hep-th/9811131}{{\tt arXiv:hep-th/9811131}}.

\bibitem{DV}
R.~Dijkgraaf and C.~Vafa, ``{Matrix models, topological strings, and
  supersymmetric gauge theories},''
  \href{http://dx.doi.org/10.1016/S0550-3213(02)00766-6}{{\em Nucl.Phys.} {\bf
  B644} (2002)  3--20},
\href{http://arxiv.org/abs/hep-th/0206255}{{\tt arXiv:hep-th/0206255
  [hep-th]}}.

\bibitem{IH}
M.~Aganagic, R.~Dijkgraaf, A.~Klemm, M.~Marino, and C.~Vafa, ``Topological
  strings and integrable hierarchies,''
  \href{http://dx.doi.org/10.1007/s00220-005-1448-9}{{\em Commun.Math.Phys.}
  {\bf 261} (2006)  451--516},
\href{http://arxiv.org/abs/hep-th/0312085}{{\tt arXiv:hep-th/0312085
  [hep-th]}}.

\bibitem{DVt}
R.~Dijkgraaf and C.~Vafa, ``{Toda Theories, Matrix Models, Topological Strings,
  and N=2 Gauge Systems},''
\href{http://arxiv.org/abs/0909.2453}{{\tt arXiv:0909.2453 [hep-th]}}.

\bibitem{toappear}
M.~Aganagic and S.~Shakirov, ``{Gauge/Vortex Duality and AGT},'' {\em to appear
  in a special volume edited by J. Teschner}  .

\bibitem{Nekrasovtalk}
N.~Nekrasov, ``{On the BPS/CFT correspondence},'' {\em Seminar at the
  University of Amsterdam} (2004)  .

\bibitem{Carlsson:2013jka}
E.~Carlsson, N.~Nekrasov, and A.~Okounkov, ``{Five dimensional gauge theories
  and vertex operators},''
\href{http://arxiv.org/abs/1308.2465}{{\tt arXiv:1308.2465 [math.RT]}}.

\bibitem{Benini:2009gi}
F.~Benini, S.~Benvenuti, and Y.~Tachikawa, ``{Webs of five-branes and N=2
  superconformal field theories},''
  \href{http://dx.doi.org/10.1088/1126-6708/2009/09/052}{{\em JHEP} {\bf 0909}
  (2009)  052},
\href{http://arxiv.org/abs/0906.0359}{{\tt arXiv:0906.0359 [hep-th]}}.

\bibitem{T1}
L.~Bao, V.~Mitev, E.~Pomoni, M.~Taki, and F.~Yagi, ``{Non-Lagrangian Theories
  from Brane Junctions},''
\href{http://arxiv.org/abs/1310.3841}{{\tt arXiv:1310.3841 [hep-th]}}.

\bibitem{T2}
H.~Hayashi, H.-C. Kim, and T.~Nishinaka, ``{Topological strings and 5d TN
  partition functions},''
\href{http://arxiv.org/abs/1310.3854}{{\tt arXiv:1310.3854 [hep-th]}}.

\bibitem{T3}
O.~Bergman, D.~Rodriguez-Gomez, and G.~Zafrir, ``{5-Brane Webs, Symmetry
  Enhancement, and Duality in 5d Supersymmetric Gauge Theory},''
\href{http://arxiv.org/abs/1311.4199}{{\tt arXiv:1311.4199 [hep-th]}}.

\bibitem{NP}
N.~Nekrasov and V.~Pestun, ``{Seiberg-Witten geometry of four dimensional N=2
  quiver gauge theories},''
\href{http://arxiv.org/abs/1211.2240}{{\tt arXiv:1211.2240 [hep-th]}}.

\bibitem{Nekrasov:2013xda}
N.~Nekrasov, V.~Pestun, and S.~Shatashvili, ``{Quantum geometry and quiver
  gauge theories},''
\href{http://arxiv.org/abs/1312.6689}{{\tt arXiv:1312.6689 [hep-th]}}.

\bibitem{NHS}
H.~Nakajima, ``{Handsaw quiver varieties and finite W-algebras},''{\em ArXiv
  e-prints} (July, 2011)  , \href{http://arxiv.org/abs/1107.5073}{{\tt
  arXiv:1107.5073 [math.QA]}}.

\bibitem{Witten_anewlook}
E.~Witten, ``A new look at the path integral of quantum mechanics,'' in {\em
  Surveys in differential geometry. {V}olume {XV}. {P}erspectives in
  mathematics and physics}, vol.~15 of {\em Surv. Differ. Geom.}, pp.~345--419.
\newblock Int. Press, Somerville, MA, 2011.
\newblock \href{http://arxiv.org/abs/hep-th/1009.6032}{{\tt
  arXiv:hep-th/1009.6032}}.

\bibitem{TL}
J.~Teschner, ``{Liouville theory revisited},''
  \href{http://dx.doi.org/10.1088/0264-9381/18/23/201}{{\em Class.Quant.Grav.}
  {\bf 18} (2001)  R153--R222},
\href{http://arxiv.org/abs/hep-th/0104158}{{\tt arXiv:hep-th/0104158
  [hep-th]}}.

\bibitem{Lawrence:1997jr}
A.~E. Lawrence and N.~Nekrasov, ``{Instanton sums and five-dimensional gauge
  theories},'' \href{http://dx.doi.org/10.1016/S0550-3213(97)00694-9}{{\em
  Nucl.Phys.} {\bf B513} (1998)  239--265},
\href{http://arxiv.org/abs/hep-th/9706025}{{\tt arXiv:hep-th/9706025
  [hep-th]}}.

\bibitem{N2}
N.~A. Nekrasov, ``{Seiberg-Witten prepotential from instanton counting},'' {\em
  Adv.Theor.Math.Phys.} {\bf 7} (2004)  831--864,
\href{http://arxiv.org/abs/hep-th/0206161}{{\tt arXiv:hep-th/0206161
  [hep-th]}}.

\bibitem{NO}
N.~Nekrasov and A.~Okounkov, ``{Seiberg-Witten theory and random partitions},''
\href{http://arxiv.org/abs/hep-th/0306238}{{\tt arXiv:hep-th/0306238
  [hep-th]}}.

\bibitem{W7}
E.~Witten, ``{Solutions of four-dimensional field theories via M theory},''
  \href{http://dx.doi.org/10.1016/S0550-3213(97)00416-1}{{\em Nucl.Phys.} {\bf
  B500} (1997)  3--42},
\href{http://arxiv.org/abs/hep-th/9703166}{{\tt arXiv:hep-th/9703166
  [hep-th]}}.

\bibitem{Katz:1997eq}
S.~Katz, P.~Mayr, and C.~Vafa, ``{Mirror symmetry and exact solution of 4-D N=2
  gauge theories: 1.},'' {\em Adv.Theor.Math.Phys.} {\bf 1} (1998)  53--114,
\href{http://arxiv.org/abs/hep-th/9706110}{{\tt arXiv:hep-th/9706110
  [hep-th]}}.

\bibitem{Mironov:2012uh}
A.~Mironov, A.~Morozov, Y.~Zenkevich, and A.~Zotov, ``{Spectral Duality in
  Integrable Systems from AGT Conjecture},''
  \href{http://dx.doi.org/10.1134/S0021364013010062}{{\em JETP Lett.} {\bf 97}
  (2013)  45--51},
\href{http://arxiv.org/abs/1204.0913}{{\tt arXiv:1204.0913 [hep-th]}}.

\bibitem{Mironov:2013xva}
A.~Mironov, A.~Morozov, B.~Runov, Y.~Zenkevich, and A.~Zotov, ``{Spectral
  dualities in XXZ spin chains and five dimensional gauge theories},''
\href{http://arxiv.org/abs/1307.1502}{{\tt arXiv:1307.1502 [hep-th]}}.

\bibitem{Nekrasov:2004vw}
N.~Nekrasov and S.~Shadchin, ``{ABCD of instantons},''
  \href{http://dx.doi.org/10.1007/s00220-004-1189-1}{{\em Commun.Math.Phys.}
  {\bf 252} (2004)  359--391},
\href{http://arxiv.org/abs/hep-th/0404225}{{\tt arXiv:hep-th/0404225
  [hep-th]}}.

\bibitem{V:3}
Y.~Tachikawa, ``{Instantons and W-algebras},'' \href{http://arxiv.org/abs/to
  appear}{{\tt to appear}}.

\bibitem{Awata:2008ed}
H.~Awata and H.~Kanno, ``{Refined BPS state counting from Nekrasov's formula
  and Macdonald functions},''
  \href{http://dx.doi.org/10.1142/S0217751X09043006}{{\em Int.J.Mod.Phys.} {\bf
  A24} (2009)  2253--2306},
\href{http://arxiv.org/abs/0805.0191}{{\tt arXiv:0805.0191 [hep-th]}}.

\bibitem{Leung:1997tw}
N.~C. Leung and C.~Vafa, ``{Branes and toric geometry},'' {\em
  Adv.Theor.Math.Phys.} {\bf 2} (1998)  91--118,
\href{http://arxiv.org/abs/hep-th/9711013}{{\tt arXiv:hep-th/9711013
  [hep-th]}}.

\bibitem{Aharony:1997bh}
O.~Aharony, A.~Hanany, and B.~Kol, ``{Webs of (p,q) five-branes,
  five-dimensional field theories and grid diagrams},''
  \href{http://dx.doi.org/10.1088/1126-6708/1998/01/002}{{\em JHEP} {\bf 9801}
  (1998)  002},
\href{http://arxiv.org/abs/hep-th/9710116}{{\tt arXiv:hep-th/9710116
  [hep-th]}}.

\bibitem{NS}
N.~A. Nekrasov and S.~L. Shatashvili, ``{Supersymmetric vacua and Bethe
  ansatz},'' \href{http://dx.doi.org/10.1016/j.nuclphysbps.2009.07.047}{{\em
  Nucl. Phys. Proc. Suppl.} {\bf 192-193} (2009)  91--112},
\href{http://arxiv.org/abs/0901.4744}{{\tt arXiv:0901.4744 [hep-th]}}.

\bibitem{NS1}
N.~A. Nekrasov and S.~L. Shatashvili, ``{Quantum integrability and
  supersymmetric vacua},'' \href{http://dx.doi.org/10.1143/PTPS.177.105}{{\em
  Prog.Theor.Phys.Suppl.} {\bf 177} (2009)  105--119},
\href{http://arxiv.org/abs/0901.4748}{{\tt arXiv:0901.4748 [hep-th]}}.

\bibitem{OV}
H.~Ooguri and C.~Vafa, ``{Knot invariants and topological strings},''
  \href{http://dx.doi.org/10.1016/S0550-3213(00)00118-8}{{\em Nucl.Phys.} {\bf
  B577} (2000)  419--438},
\href{http://arxiv.org/abs/hep-th/9912123}{{\tt arXiv:hep-th/9912123
  [hep-th]}}.

\bibitem{Douglas:1996sw}
M.~R. Douglas and G.~W. Moore, ``{D-branes, quivers, and ALE instantons},''
\href{http://arxiv.org/abs/hep-th/9603167}{{\tt arXiv:hep-th/9603167
  [hep-th]}}.

\bibitem{Cachazo:2001gh}
F.~Cachazo, S.~Katz, and C.~Vafa, ``{Geometric transitions and N=1 quiver
  theories},''
\href{http://arxiv.org/abs/hep-th/0108120}{{\tt arXiv:hep-th/0108120
  [hep-th]}}.

\bibitem{MV}
M.~Aganagic and C.~Vafa, ``Mirror symmetry, D-branes and counting holomorphic
  discs,''
\href{http://arxiv.org/abs/hep-th/0012041}{{\tt arXiv:hep-th/0012041
  [hep-th]}}.

\bibitem{Witten:1997ep}
E.~Witten, ``{Branes and the dynamics of QCD},''
  \href{http://dx.doi.org/10.1016/S0550-3213(97)00648-2}{{\em Nucl.Phys.} {\bf
  B507} (1997)  658--690},
\href{http://arxiv.org/abs/hep-th/9706109}{{\tt arXiv:hep-th/9706109
  [hep-th]}}.

\bibitem{Strominger:1995cz}
A.~Strominger, ``{Massless black holes and conifolds in string theory},''
  \href{http://dx.doi.org/10.1016/0550-3213(95)00287-3}{{\em Nucl.Phys.} {\bf
  B451} (1995)  96--108},
\href{http://arxiv.org/abs/hep-th/9504090}{{\tt arXiv:hep-th/9504090
  [hep-th]}}.

\bibitem{Greene:1995h}
B.~R. Greene, D.~R. Morrison, and A.~Strominger, ``{Black hole condensation and
  the unification of string vacua},''
  \href{http://dx.doi.org/10.1016/0550-3213(95)00371-X}{{\em Nucl.Phys.} {\bf
  B451} (1995)  109--120},
\href{http://arxiv.org/abs/hep-th/9504145}{{\tt arXiv:hep-th/9504145
  [hep-th]}}.

\bibitem{Greene:1996dh}
B.~R. Greene, D.~R. Morrison, and C.~Vafa, ``{A Geometric realization of
  confinement},'' \href{http://dx.doi.org/10.1016/S0550-3213(96)00465-8}{{\em
  Nucl.Phys.} {\bf B481} (1996)  513--538},
\href{http://arxiv.org/abs/hep-th/9608039}{{\tt arXiv:hep-th/9608039
  [hep-th]}}.

\bibitem{Hori:1997zj}
K.~Hori, H.~Ooguri, and C.~Vafa, ``{NonAbelian conifold transitions and N=4
  dualities in three-dimensions},''
  \href{http://dx.doi.org/10.1016/S0550-3213(97)00529-4}{{\em Nucl.Phys.} {\bf
  B504} (1997)  147--174},
\href{http://arxiv.org/abs/hep-th/9705220}{{\tt arXiv:hep-th/9705220
  [hep-th]}}.

\bibitem{Alday:2009fs}
L.~F. Alday, D.~Gaiotto, S.~Gukov, Y.~Tachikawa, and H.~Verlinde, ``{Loop and
  surface operators in N=2 gauge theory and Liouville modular geometry},''
  \href{http://dx.doi.org/10.1007/JHEP01(2010)113}{{\em JHEP} {\bf 1001} (2010)
   113},
\href{http://arxiv.org/abs/0909.0945}{{\tt arXiv:0909.0945 [hep-th]}}.

\bibitem{Dimofte:2010tz}
T.~Dimofte, S.~Gukov, and L.~Hollands, ``{Vortex Counting and Lagrangian
  3-manifolds},'' \href{http://dx.doi.org/10.1007/s11005-011-0531-8}{{\em Lett.
  Math. Phys.} {\bf 98} (2011)  225--287},
\href{http://arxiv.org/abs/1006.0977}{{\tt arXiv:1006.0977 [hep-th]}}.

\bibitem{YK}
H.~Kanno and Y.~Tachikawa, ``{Instanton counting with a surface operator and
  the chain-saw quiver},''
  \href{http://dx.doi.org/10.1007/JHEP06(2011)119}{{\em JHEP} {\bf 1106} (2011)
   119},
\href{http://arxiv.org/abs/1105.0357}{{\tt arXiv:1105.0357 [hep-th]}}.

\bibitem{Wittenphases}
E.~Witten, ``{Phases of N=2 theories in two-dimensions},''
  \href{http://dx.doi.org/10.1016/0550-3213(93)90033-L}{{\em Nucl.Phys.} {\bf
  B403} (1993)  159--222},
\href{http://arxiv.org/abs/hep-th/9301042}{{\tt arXiv:hep-th/9301042
  [hep-th]}}.

\bibitem{Donagi:2007hi}
R.~Donagi and E.~Sharpe, ``{GLSM's for partial flag manifolds},''
  \href{http://dx.doi.org/10.1016/j.geomphys.2008.07.010}{{\em J.Geom.Phys.}
  {\bf 58} (2008)  1662--1692},
\href{http://arxiv.org/abs/0704.1761}{{\tt arXiv:0704.1761 [hep-th]}}.

\bibitem{AS}
M.~Aganagic and S.~Shakirov, ``{Knot Homology from Refined Chern-Simons
  Theory},''
\href{http://arxiv.org/abs/1105.5117}{{\tt arXiv:1105.5117 [hep-th]}}.

\bibitem{AS2}
M.~Aganagic and S.~Shakirov, ``{Refined Chern-Simons Theory and Knot
  Homology},''
\href{http://arxiv.org/abs/1202.2489}{{\tt arXiv:1202.2489 [hep-th]}}.

\bibitem{tt1}
S.~Cecotti, D.~Gaiotto, and C.~Vafa, ``{tt* Geometry in 3 and 4 Dimensions},''
\href{http://arxiv.org/abs/1312.1008}{{\tt arXiv:1312.1008 [hep-th]}}.

\bibitem{tt2}
C.~Vafa, ``{tt* Geometry and a Twistorial Extension of Topological Strings},''
\href{http://arxiv.org/abs/1402.2674}{{\tt arXiv:1402.2674 [hep-th]}}.

\bibitem{Fateev:2007ab}
V.~Fateev and A.~Litvinov, ``{Correlation functions in conformal Toda field
  theory. I.},'' \href{http://dx.doi.org/10.1088/1126-6708/2007/11/002}{{\em
  JHEP} {\bf 0711} (2007)  002},
\href{http://arxiv.org/abs/0709.3806}{{\tt arXiv:0709.3806 [hep-th]}}.

\bibitem{Fateev:1987zh}
V.~Fateev and S.~L. Lukyanov, ``{The Models of Two-Dimensional Conformal
  Quantum Field Theory with Z(n) Symmetry},''
\href{http://dx.doi.org/10.1142/S0217751X88000205}{{\em Int.J.Mod.Phys.} {\bf
  A3} (1988)  507}.

\bibitem{FR1}
E.~{Frenkel} and N.~{Reshetikhin}, ``{Deformations of W-algebras associated to
  simple Lie algebras},'' in {\em eprint arXiv:q-alg/9708006}, p.~8006.
\newblock Aug., 1997.

\bibitem{NS2}
N.~A. Nekrasov and S.~L. Shatashvili, ``{Quantization of Integrable Systems and
  Four Dimensional Gauge Theories},''
\href{http://arxiv.org/abs/0908.4052}{{\tt arXiv:0908.4052 [hep-th]}}.

\bibitem{WittenGukov}
S.~Gukov and E.~Witten, ``{Branes and Quantization},'' {\em
  Adv.Theor.Math.Phys.} {\bf 13} (2009)  ,
\href{http://arxiv.org/abs/0809.0305}{{\tt arXiv:0809.0305 [hep-th]}}.

\bibitem{NW}
N.~Nekrasov and E.~Witten, ``{The Omega Deformation, Branes, Integrability, and
  Liouville Theory},'' \href{http://dx.doi.org/10.1007/JHEP09(2010)092}{{\em
  JHEP} {\bf 1009} (2010)  092},
\href{http://arxiv.org/abs/1002.0888}{{\tt arXiv:1002.0888 [hep-th]}}.

\bibitem{Hanany:1996ie}
A.~Hanany and E.~Witten, ``{Type IIB superstrings, BPS monopoles, and
  three-dimensional gauge dynamics},''
  \href{http://dx.doi.org/10.1016/S0550-3213(97)00157-0}{{\em Nucl.Phys.} {\bf
  B492} (1997)  152--190},
\href{http://arxiv.org/abs/hep-th/9611230}{{\tt arXiv:hep-th/9611230
  [hep-th]}}.

\end{thebibliography}\endgroup
\end{document}